\documentclass[aps,pra,reprint,amsmath,amssymb,showpacs,floatfix]{revtex4-1}

\usepackage{graphicx}

\newcommand{\ket}[1]{\ensuremath{\vert{#1\rangle}}} 
\newcommand{\bra}[1]{\ensuremath{{\langle #1}\vert}}
\newcommand{\braket}[2]{\ensuremath{{\langle #1}\vert{#2 \rangle}}}
\newcommand{\op}[1]{\hat{#1}}
\newcommand{\D}{\text{d}}
\newcommand{\I}{\text{i}}
\newcommand{\E}{\text{e}}
\providecommand{\abs}[1]{\left\lvert#1\right\rvert}

\usepackage[colorlinks,citecolor=blue,urlcolor=blue,linkcolor=blue,hyperindex,breaklinks]{hyperref}

\begin{document}

\title{State disturbance and pointer shift in protective quantum measurements}
\author{Maximilian Schlosshauer}
\affiliation{\small Department of Physics, University of Portland, 5000 North Willamette Boulevard, Portland, Oregon 97203, USA}

\begin{abstract}
We investigate the disturbance of the state of a quantum system in a protective measurement for finite measurement times and different choices of the time-dependent system--apparatus coupling function. The ability to minimize this state disturbance is essential to protective measurement. We show that for a coupling strength that remains constant during the measurement interaction of duration $T$, the state disturbance scales as $T^{-2}$, while a simple smoothing of the coupling function significantly improves the scaling behavior to $T^{-6}$. We also prove that the shift of the apparatus pointer in the course of a protective measurement is independent of the particular time dependence of the coupling function, suggesting that the guiding principle for choosing the coupling function should be the minimization of the state disturbance. Our results illuminate the dynamics of protective measurement under realistic circumstances and may aid in the experimental realization of such measurements.\\[-.1cm]

\noindent Journal reference: \emph{Phys.\ Rev.\ A\ }\textbf{90}, 052106 (2014)
\end{abstract}

\pacs{03.65.Ta, 03.65.Wj}

\maketitle

\section{Introduction}

Protective measurement \cite{Aharonov:1993:qa,Aharonov:1993:jm,Aharonov:1996:fp,Dass:1999:az,Vaidman:2009:po} is a quantum measurement scheme in which an apparatus is weakly coupled to a quantum system for an extended period of time. If the system starts out in a nondegenerate eigenstate of its Hamiltonian and the interaction is sufficiently weak and long, then expectation values of observables of the system can be measured without appreciably disturbing the state of the system. Since measurement of a sufficient number of expectation values allows one to reconstruct a quantum state, protective measurement, if suitably implemented (see Ref.~\cite{Dass:1999:az} for a discussion of constraints and complications), may enable reconstruction of the quantum state of an individual system. This provides a perspective on state reconstruction different from that associated with conventional ensemble state tomography based on strong \cite{Vogel:1989:uu,Smithey:1993:lm,Breitenbach:az,White:1999:az} or weak \cite{Aharonov:1988:mz,Lundeen:2011:ii,Lundeen:2012:rr} measurements. 

Only an infinitely weak or infinitely slowly changing measurement interaction will not disturb the state of a protectively measured system; this follows directly from perturbation theory and the quantum adiabatic theorem \cite{Born:1928:yf}. Outside these limiting cases, however, protective measurement, if it is to yield new information, cannot avoid disturbing the state of the system, in agreement with general results concerning the fundamental tradeoff between quantum state disturbance and information gain \cite{Fuchs:1996:op} and the independence of the maximum possible information gain in a quantum measurement from the method of measurement \cite{Ariano:1996:om}. From a fundamental point of view, this inevitable state disturbance disproves suggestions \cite{Aharonov:1993:qa,Aharonov:1993:jm,Gao:2013:om} that protective measurement permits state measurement akin to a classical state and bears on the meaning of the wavefunction (see Refs.~\cite{Alter:1997:oo,Dass:1999:az,Schlosshauer:2014:tp} for discussions of this important foundational point).

This limitation, however, does not invalidate the potential practical usefulness of protective measurement.  Implementation of protective measurement would be interesting and important both from a fundamental point of view (as the realization of a new quantum measurement scheme) and from a practical point of view (enabling quantum state tomography for single systems). Just like traditional ensemble quantum state tomography, protective measurement provides a way of (approximately) reconstructing a quantum state. The fidelity of any such reconstruction can be measured in terms of the disturbance of the initial state of the system incurred during the measurement. At the heart of protective measurement is the idea that this state disturbance can be made arbitrarily small, such that repeated measurements on the same system permit reconstruction of its initial state with arbitrarily high fidelity \cite{Aharonov:1993:qa,Aharonov:1993:jm,Aharonov:1996:fp,Dass:1999:az}. Therefore, for practical implementations of protective measurement it is essential to gain a precise and quantitative understanding of how one may reduce the state disturbance incurred during a protective measurement while simultaneously maintaining appreciable information gain. 

Despite its significance, however, the problem of state disturbance in protective measurement has not yet been adequately studied. Instead, the existing literature (see, e.g., Refs.~\cite{Aharonov:1993:qa,Aharonov:1993:jm,Aharonov:1996:fp,Vaidman:2009:po}) has relied on the consideration of mathematical limits involving infinitely long, infinitely weak, and/or infinitely slowly changing (adiabatic) measurement interactions, for which the state of the system can be shown to remain unchanged during the measurement. This, however, leaves open the important question of precisely how much the initial state will be disturbed in the physically relevant case of finite measurement times and interaction strengths, and how this disturbance depends on the particular choice of the coupling function describing the time dependence of the system--apparatus interaction.

This paper addresses this question. We study the state disturbance in a protective measurement for different coupling functions and make precise the dependence of the state disturbance on the physical parameters of the system and the measurement interaction. In particular, we show how a careful choice of the coupling function can dramatically reduce the state disturbance. In turn, this raises the question of whether and how the information gain during the measurement, represented by the shift of the apparatus pointer to a position indicating the expectation value of the measured observable of the system, depends on the particular choice of the coupling function. We show that, to a good approximation, the shift of the apparatus pointer is in fact independent of the choice of the coupling function (under the customary assumption that the coupling function is appropriately normalized). 
 
This paper is organized as follows. In Sec.~\ref{sec:prot-meas} we introduce the basic concepts of protective measurement and develop a framework, based on time-dependent perturbation theory, for describing the dynamics of protective measurement for arbitrary time-dependent system--apparatus coupling strengths, finite measurement times, and up to any order in the interaction. In Sec.~\ref{sec:state-disturbance} we investigate the disturbance of the initial state in the course of a protective measurement for several different coupling functions. In particular, we investigate how this disturbance depends on the time dependence of the coupling and the duration of the measurement. In Sec.~\ref{sec:pointer-shift} we derive an expression for the pointer shift for arbitrary time-dependent coupling functions and show that it is generic. We discuss our results in Sec.~\ref{sec:discussion}. In Appendix~\ref{sec:high-order-corr} we investigate the influence of higher-order perturbative corrections.

\section{\label{sec:prot-meas}Protective measurement}

Following the standard framework for protective measurement (see, e.g., Refs.~\cite{Aharonov:1993:qa,Aharonov:1993:jm,Aharonov:1996:fp,Dass:1999:az}), we consider a system $S$ and apparatus $A$ with time-independent self-Hamiltonians $\op{H}_S$ and $\op{H}_A$, respectively, where the spectrum of $\op{H}_S$ is assumed to be nondegenerate. We let the system and apparatus interact via the time-dependent interaction Hamiltonian 
\begin{equation}\label{eq:lalaa}
\op{H}_\text{int}(t) = g(t)\op{O} \otimes \op{P}.
\end{equation}
Here, $g(t)$ is a non-negative function representing a time-dependent coupling strength. We let the interaction start at $t=-T/2$ and conclude at $t=T/2$, so $g(t)= 0$ outside the interval $[-T/2,T/2]$ and the total measurement time is $T$. We also normalize $g(t)$ according to 
\begin{equation}\label{eq:normal}
\int_{-T/2}^{T/2} \D t\, g(t) =1. 
\end{equation}
The normalization effectively links the interaction strength to the duration of the interaction $T$: The larger $T$ is, the weaker the average strength of the interaction. The system observable $\op{O}$ can be freely chosen and need not commute with $\op{H}_S$. The operator $\op{P}$ denotes the momentum conjugate of the pointer variable $\op{X}$ of the apparatus. In what follows we adopt the customary assumption that $\op{P}$ commutes with $\op{H}_A$, i.e., that the interaction Hamiltonian is diagonal in the energy eigenbasis of the system. While this assumption is not necessary for a protective measurement to obtain \cite{Dass:1999:az}, it simplifies the subsequent calculations and is innocent in the context of the present paper, which focuses on the effect of the measurement on the subspace of the system. In this case, the operator $\op{P}$ is a constant of motion of the total Hamiltonian $\op{H}(t)=\op{H}_S+\op{H}_A+\op{H}_\text{int}(t)$ and there exists a set of simultaneous orthonormal eigenstates $\{\ket{A_i}\}$ of $\op{P}$ and $\op{H}_A$ with spectra $\{a_i\}$ and $\{\epsilon_i\}$ obeying 
\begin{equation}\label{eq:haaz}
\op{P}\ket{A_i}=a_i\ket{A_i}, \qquad \op{H}_A\ket{A_i}=\epsilon_i\ket{A_i}.
\end{equation}

In a standard impulsive (strong) measurement \cite{vonNeumann:1955:ii}, the interaction time $T$ is short and thus $g(t)$ is large for $t \in [-T/2,T/2]$. In this case, the evolution is dominated by the interaction Hamiltonian, leading to strong entanglement between the system and apparatus and therefore to a significant disturbance of the initial state of the system by the measurement. The idea of protective measurement is to minimize the state disturbance, and yet to obtain meaningful information about the system in a single measurement, by making the interaction weak while leaving it turned on for a long time $T$, so as to ensure an appreciable shift of the apparatus pointer. This long-interaction limit is in contrast with the weak-measurement protocol \cite{Aharonov:1988:mz}, where the measurement interaction is both weak and short and the insignificant pointer shift arising from a single measurement is compensated for by repeating the measurement on a large ensemble of systems.

A protective measurement proceeds from the assumption that the system $S$ starts (at $t=-T/2$) in a nondegenerate eigenstate $\ket{n}$ of $\op{H}_S$ with eigenvalue $E_n$. The initial composite state of system and apparatus is taken to be the product state
\begin{equation}\label{eq:1fbhjsbfkj}
\ket{\Psi(-T/2)} = \ket{n}\ket{\phi(x_0)} = \ket{n} \sum_i \braket{A_i}{\phi(x_0)} \ket{A_i}.
\end{equation}
Here $\ket{\phi(x_0)}$ is a Gaussian wave packet of eigenstates of the pointer variable $\op{X}$ centered on $x_0$, representing the premeasurement ready state of the apparatus pointer, where the momentum of the pointer is assumed to be bounded. What we are interested in is the final composite state $\ket{\Psi(T/2)}$ at the conclusion of the measurement interaction at time $t=T/2$. We will now discuss this problem, first by reviewing the case of constant $g(t)=1/T$ for $t \in [-T/2,T/2]$ and then by developing the solution for arbitrary time dependent $g(t)$.

\subsection{Constant coupling}

Existing studies of protective measurement (see, e.g., Refs.~\cite{Aharonov:1993:qa,Aharonov:1993:jm,Aharonov:1996:fp,Dass:1999:az,Vaidman:2009:po,Gao:2013:om}) have considered two limiting scenarios. In the first scenario, the measurement interaction is assumed to be turned on and off infinitely slowly such that, according to the quantum adiabatic theorem \cite{Born:1928:yf}, the system remains in an eigenstate of the time-dependent Hamiltonian at all times. Since at the end of the interaction the Hamiltonian returns to its pre-measurement form, i.e., $\op{H}(T/2)=\op{H}(-T/2)$, the system will also deterministically return to its initial state. In the second scenario, which is the one that is typically used in expositions of protective measurement (see, e.g., Refs.~\cite{Dass:1999:az,Vaidman:2009:po}), the interaction is discontinuously turned on at the initial time $t=-T/2$, kept constant at strength $1/T$ during the duration $T$ of the measurement, and then discontinuously turned off again at $t=T/2$. This corresponds to the choice
\begin{equation}\label{eq:igv34763578}
g(t) = \begin{cases} 1/T, & -T/2 \le t \le T/2 \\ 0 & \text{otherwise}. \end{cases}
\end{equation}
In the following we will refer to this choice of $g(t)$ as constant coupling. The total Hamiltonian $\op{H}$ is now effectively time independent, which allows the problem to be treated using time-independent perturbation theory. 

We briefly review the derivation (see, e.g., Ref.~\cite{Dass:1999:az} for details). We consider $\op{H}_\text{int}=(1/T)\op{O} \otimes \op{P}$ as a (small) time-independent perturbation to $\op{H}_0=\op{H}_S+\op{H}_A$. Starting from the initial state $\ket{\Psi(t=-T/2)}=\ket{n}\ket{\phi(x_0)}$ [see Eq.~\eqref{eq:1fbhjsbfkj}], the final composite state at the conclusion of the measurement interaction at time $T$ is
\begin{align}
\ket{\Psi(t=T/2)} &=\E^{-\I \op{H} T/\hbar} \ket{n}\ket{\phi(x_0)} \notag \\
&= \sum_{m,i} \E^{-\I \widetilde{E} (m,a_i) T/\hbar} \braket{E_m^S(a_i)}{n} \notag \\ &\quad \times \braket{A_i}{\phi(x_0)} \ket{E_m^S(a_i)}\ket{A_i},\label{eq:jkbknjd}
\end{align}
where $\ket{E_m^S(a_i)}$ are the eigenstates of the system-dependent part of $\op{H}$ defined by $\op{H}'_S(a_i) = \op{H}_S+\frac{1}{T} (a_i \op{O})$ and 
\begin{align}\label{eq:fnsnlk1}
\widetilde{E}(m,a_i) &= \bra{E_m^S(a_i)} \op{H}_S \ket{E_m^S(a_i)} + \epsilon_i \notag \\ &\quad + \frac{1}{T}a_i \bra{E_m^S(a_i)}\op{O} \ket{E_m^S(a_i)}
\end{align}
are the eigenvalues of the eigenstates $\ket{E_m^S(a_i)}\ket{A_i}$ of $\op{H}$. Using zeroth-order time-independent perturbation theory corresponding to the limit $T\rightarrow\infty$, one expands the eigenvalues $\widetilde{E} (m,a_i)$ to first order in $1/T$ (such that the argument of the exponential $\E^{-\I \widetilde{E} (m,a_i) T/\hbar}$ is of zeroth order in $1/T$),
\begin{equation}\label{eq:fnsnlk11}
\widetilde{E} (m,a_i) \approx  E_m + \epsilon_i + \frac{1}{T}a_i \bra{m}\op{O} \ket{m},
\end{equation}
and replaces the exact eigenstates $\ket{E_m^S(a_i)}$ by their zeroth-order approximations $\ket{m}$. Reintroducing the operators $\op{H}_A$ and $\op{P}$ in the exponent of the time-evolution operator, the final zeroth-order system--apparatus state is therefore
\begin{align}\label{eq:isvgaazxxx}
\ket{\Psi^{(0)}(t=T/2)} &= \E^{-\I E_nT/\hbar} \ket{n} \E^{-\I \op{H}_A T/\hbar}\notag \\ &\quad \times \E^{-\I \op{P} \bra{n} \op{O} \ket{n}/\hbar} \ket{\phi(x_0)}. 
\end{align}
The operator $\E^{-\I \op{P} \bra{n} \op{O} \ket{n}/\hbar}$ shifts the center of the wave packet $\ket{\phi(x_0)}$ by an amount equal to $\bra{n} \op{O} \ket{n}$. In this way, information about the expectation value of $\op{O}$ in the initial state $\ket{n}$ becomes encoded in the pointer position and the final composite state, to zeroth order, is
\begin{align}\label{eq:gctctwg}
\ket{\Psi^{(0)}(t=T/2)} = \E^{-\I E_nT/\hbar} \ket{n} \E^{-\I \op{H}_A T/\hbar}\ket{\phi(x_0+\langle \op{O} \rangle_n)}.
\end{align}
In this strict limit $T \rightarrow \infty$ (corresponding to an infinitely weak interaction), the initial state of the system remains unchanged and there is no entanglement between the system and apparatus. It is in this sense that the state of the system is protected. The protection is provided by the dominant Hamiltonian $\op{H}_S$ such that the interaction Hamiltonian $\op{H}_\text{int}$ can be treated as a small perturbation whose effect on the system is negligible in the limit $T \rightarrow \infty$, even though it still induces a finite pointer shift in the apparatus. Note that once the state is appropriately protected, information about the expectation value of \emph{any} observable $\op{O}$ can be obtained. This permits, at least in principle, sequential protective measurements of many different observables using the same protection potential.

\subsection{Time-dependent coupling}

Clearly, the limit $T \rightarrow \infty$ is not physically realizable and it is therefore important to understand and explore protective measurement in the practically relevant case of finite $T$. Furthermore, rather than being restricted to the constant coupling given by Eq.~\eqref{eq:igv34763578}, we would like to consider arbitrary time-dependent coupling functions $g(t)$. To this end, we will now treat the evolution of the initial state $\ket{\Psi(-T/2)} = \ket{n}\ket{\phi(x_0)}$ [Eq.~\eqref{eq:1fbhjsbfkj}] for arbitrary $g(t)$ by using time-dependent perturbation theory, regarding $\op{H}_\text{int}(t)=g(t)\op{O} \otimes \op{P}$ as a time-dependent perturbation to $\op{H}_0=\op{H}_S+\op{H}_A$. 

As before, we shall assume $[\op{P},\op{H}_A]=0$, i.e., the perturbation commutes with the unperturbed Hamiltonian in the apparatus subspace. Then the perturbation does not connect the different energy levels $\ket{A_i}$ of the apparatus. This can also be seen from considering the evolution operator in the interaction picture, which we may symbolically write as a time-ordered exponential,
\begin{align}
\op{U}_I(-T/2,T/2)&=\mathcal{T} \exp\left[-\frac{\I}{\hbar} \int_{-T/2}^{T/2} \D t\, \op{H}_{\text{int},I}(t)\right],
\end{align}
where $\mathcal{T}$ is the time-ordering operator and the subscript $I$ denotes interaction-picture quantities. Since the interaction Hamiltonian in the interaction picture is
\begin{align}
\op{H}_{\text{int},I}(t) &= g(t)\E^{\I (\op{H}_S+\op{H}_A) t/\hbar}(\op{O} \otimes \op{P})
\E^{-\I (\op{H}_S+\op{H}_A) t/\hbar} \notag\\ &= g(t) \op{O}_I(t) \otimes\op{P},
\end{align}
the evolution operator becomes
\begin{multline}
\op{U}_I(-T/2,T/2)\\ =\mathcal{T} \exp\left[-\frac{\I}{\hbar} \left(\int_{-T/2}^{T/2} \D t\, g(t) \op{O}_I(t)\right) \otimes \op{P}\right],
\end{multline}
which is diagonal in the energy eigenbasis $\{\ket{A_i}\}$ of the apparatus.

The final composite state at the conclusion of the measurement may then be written as
\begin{align}\label{eq:1fbhjsbfk4554j}
\ket{\Psi(T/2)} &= \sum_m \E^{-\I (E_n+E_m) T/2\hbar} \sum_i C^{(i)}_{mn}(T) \ket{m} \E^{-\I \epsilon_i T/\hbar} \notag\\ & \quad \times \braket{A_i}{\phi(x_0)} \ket{A_i},
\end{align}
where the interaction-picture amplitude $C^{(i)}_{mn}(T)$ is given by
\begin{equation}\label{eq:fgjf1}
C^{(i)}_{mn}(T) = \bra{m} \left\{ \mathcal{T} \exp\left(-\frac{\I}{\hbar} a_i \int_{-T/2}^{T/2} \D t\, g(t) \op{O}_I(t) \right) \right\} \ket{n}.
\end{equation}
Note that we have used $T$ as the argument of $C^{(i)}_{mn}(T)$ in order to indicate that $C^{(i)}_{mn}(T)$ is the amplitude at the conclusion of a measurement interaction of duration $T$. From here on, we will drop the subscript $n$ in $C^{(i)}_{mn}(T)$, $C^{(i)}_{mn}(T)\equiv C^{(i)}_m(T)$, since we will be assuming throughout this paper that the system starts out in the state $\ket{n}$. 

We now express the amplitude $C^{(i)}_{m}(T)$ as a perturbative expansion (Dyson series),
\begin{align}\label{eq:y89ggy892}
C^{(i)}_{m}(T) &= \sum_{\ell=0}^\infty a_i^{\ell} A^{(\ell)}_{m}(T),
\end{align}
where $A^{(\ell)}_{m}(T)$ is the expression for the $\ell$th-order correction to the zeroth-order amplitude $A^{(0)}_{m}(T)=1$ \cite{Sakurai:1994:om},
\begin{align}\label{eq:g8fbvsv1}
A^{(\ell)}_{m}(T) &= \left(-\frac{\I}{\hbar}\right)^\ell \sum_{k_1,k_2,\hdots,k_{\ell-1}}
\bra{m} \op{O} \ket{k_1} \bra{k_1} \op{O} \ket{k_2} \notag \\ & \quad\times\bra{k_2} \op{O} \ket{k_3} \cdots \bra{k_{\ell-1}} \op{O} \ket{n} \notag \\ & \quad \times\int_{-T/2}^{T/2}  \D t' \, \E^{\I \omega_{mk_1} t'} g(t') \notag \\ & \quad \times \int_{-T/2}^{t'} \D t'' \,\E^{\I \omega_{k_1k_2} t''} g(t'') \cdots \notag \\ & \quad \times \int_{-T/2}^{t^{(\ell-1)}} \D t^{(\ell)} \,\E^{\I \omega_{k_{\ell-1} n} t^{(\ell)}} g(t^{(\ell)}).
\end{align}
Here we have introduced $\omega_{mn}\equiv (E_m-E_n)/\hbar$, which is the frequency (i.e., the inverse of the time scale) associated with the transition $\ket{n} \rightarrow \ket{m}$. Specifically, the first-order correction is
\begin{align}\label{eq:87gr782}
A^{(1)}_{m}(T) &= -\frac{\I}{\hbar} \bra{m} \op{O} \ket{n} \int_{-T/2}^{T/2} \D t\, \E^{\I \omega_{mn} t} g(t),
\end{align}
and the second-order correction is
\begin{align}\label{eq:7h7grhgrh771}
A^{(2)}_{m}(T) &= \left(-\frac{\I}{\hbar}\right)^2 \sum_k
\bra{m} \op{O} \ket{k} \bra{k} \op{O} \ket{n} \notag \\ & \quad \times \int_{-T/2}^{T/2} \D t \, \E^{\I \omega_{mk} t} g(t) \int_{-T/2}^{t} \D t' \,\E^{\I \omega_{kn} t'} g(t').
\end{align}
Using Eq.~\eqref{eq:y89ggy892}, we can then write the final composite state \eqref{eq:1fbhjsbfk4554j} as
\begin{align}
\ket{\Psi(T/2)} &= \sum_{m} \E^{-\I (E_n+E_m) T/2\hbar} \sum_{\ell=0}^\infty A^{(\ell)}_{m}(T)  \ket{m} \notag \\ & \quad \times \left(\sum_{i}\E^{-\I \epsilon_i T/\hbar}a_i^\ell \braket{A_i}{\phi(x_0)} \ket{A_i}\right).
\end{align}
We see that the interesting time-dependent dynamics are contained in the amplitude contributions $A^{(\ell)}_{m}(T)$, which specify how the initial state $\ket{n}$ of the system changes in the course of the protective measurement. In particular, these contributions tell us about the disturbance of this initial state, in the sense that they quantify the mixing of other states $\ket{m}$, that is to say, the probabilities of finding the system in a state $\ket{m}\not=\ket{n}$ at the conclusion of the measurement. It is the dependence of these terms $A^{(\ell)}_{m}(T)$ on the choice of $g(t)$, $T$, and the frequency parameters $\omega_{jk}$ that will be the focus of our investigation.

\section{\label{sec:state-disturbance}State disturbance}

In a protective measurement, the goal is to minimize the transition probabilities out of the initial state $\ket{n}$, in order to minimize the disturbance of this state in the course of the measurement interaction. To study this disturbance, we now quantitatively investigate the transition probabilities for $\ket{n} \rightarrow \ket{m} \not=\ket{n}$ for particular choices of $g(t)$ and explore the dependence of these probabilities on the total duration $T$ of the measurement interaction. 

As before, we let the interaction start at $t=-T/2$ and conclude at $t=T/2$, so $g(t)= 0$ for $t<-T/2$ and $t>T/2$. We let $g(t)$ be normalized according to Eq.~\eqref{eq:normal}. We also take $g(t)$ to be an even function, i.e., we let the turn-on and turnoff time dependence be symmetric with respect to $t=0$. Then the first-order transition amplitude at the conclusion of the measurement is, from Eq.~\eqref{eq:87gr782},
\begin{align}\label{eq:1ldkf000}
A^{(1)}_{m}(T) &= -\frac{\I}{\hbar} \bra{m} \op{O} \ket{n} \int_{-T/2}^{T/2} \D t\, \cos(\omega_{mn} t) g(t),
\end{align}
and thus the corresponding transition probability $\mathcal{P}^{(1)}_m(T)$ is
\begin{align}\label{eq:bddfhdfh}
\mathcal{P}^{(1)}_m(T) &= \abs{A^{(1)}_{m}(T)}^2 \notag \\ &= \frac{1}{\hbar^2} \abs{ \bra{m} \op{O} \ket{n}}^2 \abs{\int_{-T/2}^{T/2} \D t\, \cos(\omega_{mn} t) g(t)}^2.
\end{align}

Strictly speaking, $A^{(1)}_{m}(T)$ is not the full expression for the first-order transition amplitude but only the $T$-dependent part pertaining to the system; Eq.~\eqref{eq:1fbhjsbfk4554j} shows that the full expression (neglecting phase factors) is $\sum_i a_i A^{(1)}_{m}(T) \braket{A_i}{\phi(x_0)}$. The additional terms, however, pertain solely to properties of the apparatus and are independent of the measurement time $T$. Since we are chiefly interested in the dependence of the disturbance of the system on $T$, in what follows we can focus on the amplitude $A^{(1)}_{m}(T)$ and the corresponding transition probability $\mathcal{P}^{(1)}_m(T)$ as given by Eqs.~\eqref{eq:1ldkf000} and \eqref{eq:bddfhdfh}.

We would like to investigate the dependence of the  probability $\mathcal{P}^{(1)}_m(T)$ [Eq.~\eqref{eq:bddfhdfh}] on $g(t)$ and $T$, so the relevant quantity of interest is the Fourier transform of $g(t)$,
\begin{align}\label{eq:lvdgs}
\widetilde{g}(\omega_{mn}; T) &= \int_{-T/2}^{T/2} \D t\, \cos(\omega_{mn} t) g(t),
\end{align}
which is a function in frequency space, with all frequencies measured relative to the initial-state frequency $\omega_n=E_n/\hbar$. For the cases studied below, $\widetilde{g}(\omega_{mn};T)$ is a function of $\omega_{mn}T$, so $\widetilde{g}(\omega_{mn};T) \equiv \widetilde{g}(\omega_{mn}T)$. Note that $\omega_{mn}T$ is a dimensionless quantity that measures the ratio of the total measurement time $T$ to the internal time scale $\omega_{mn}^{-1}$ associated with the transition $\ket{n}\rightarrow\ket{m}$.

In a protective measurement, we would like to minimize the transition probability, which, as Eq.~\eqref{eq:bddfhdfh} shows, is proportional to $\abs{\widetilde{g}(\omega_{mn}T)}^2$. In light of the general relationship between a function and its Fourier transform, we expect that these goals can be accomplished by increasing $T$, i.e., by increasing the width of $g(t)$, and by making $g(t)$ smoother and less rapidly changing. We will now verify these intuitions for different choices of $g(t)$: constant $g(t)$ as given by Eq.~\eqref{eq:igv34763578} (Sec.~\ref{sec:time-indep-coupl}), constant $g(t)$ with a linear turn-on and turnoff (Sec.~\ref{sec:turnonoff}), and a smoothly varying $g(t)$ following a raised-cosine function (Sec.~\ref{sec:smoothly-vary-coupl}). 
  
\subsection{\label{sec:time-indep-coupl}Constant coupling}

\begin{figure*}
\begin{flushleft}
\hspace*{2.5cm} (a) \hspace{6.2cm} (b)
\end{flushleft}

\includegraphics[scale=.55]{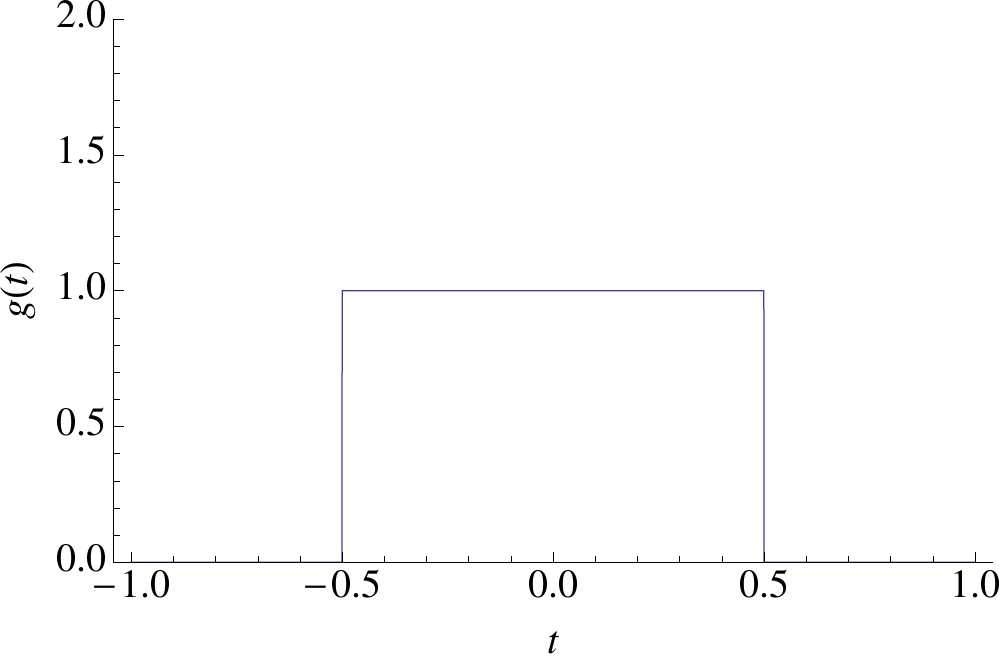}\qquad \qquad
\includegraphics[scale=.55]{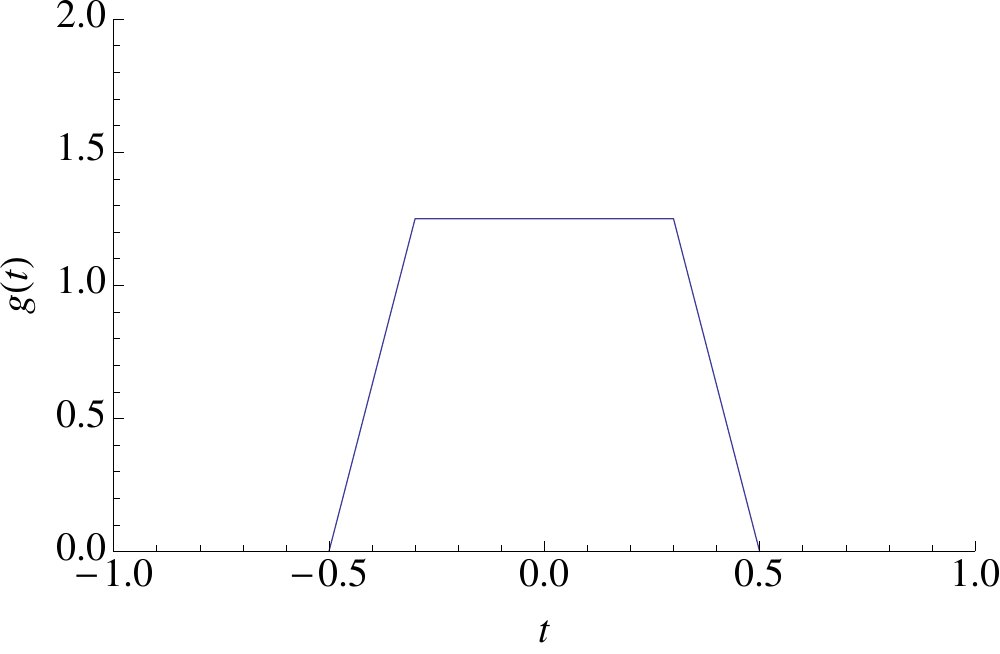}

\begin{flushleft}
\hspace*{2.5cm} (c) \hspace{6.2cm} (d)
\end{flushleft}

\includegraphics[scale=.55]{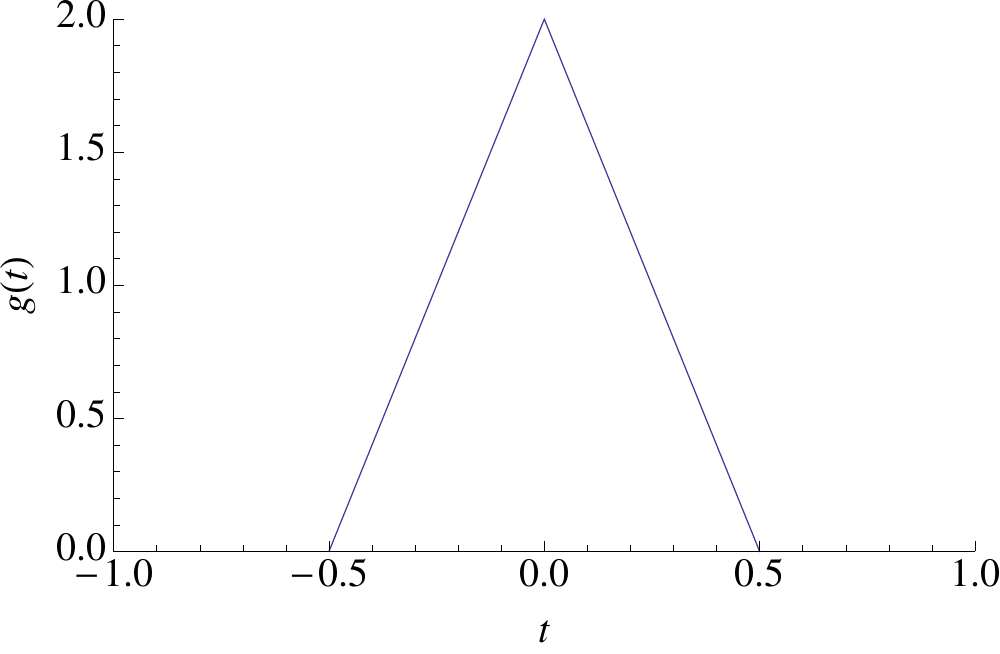}\qquad \qquad
\includegraphics[scale=.55]{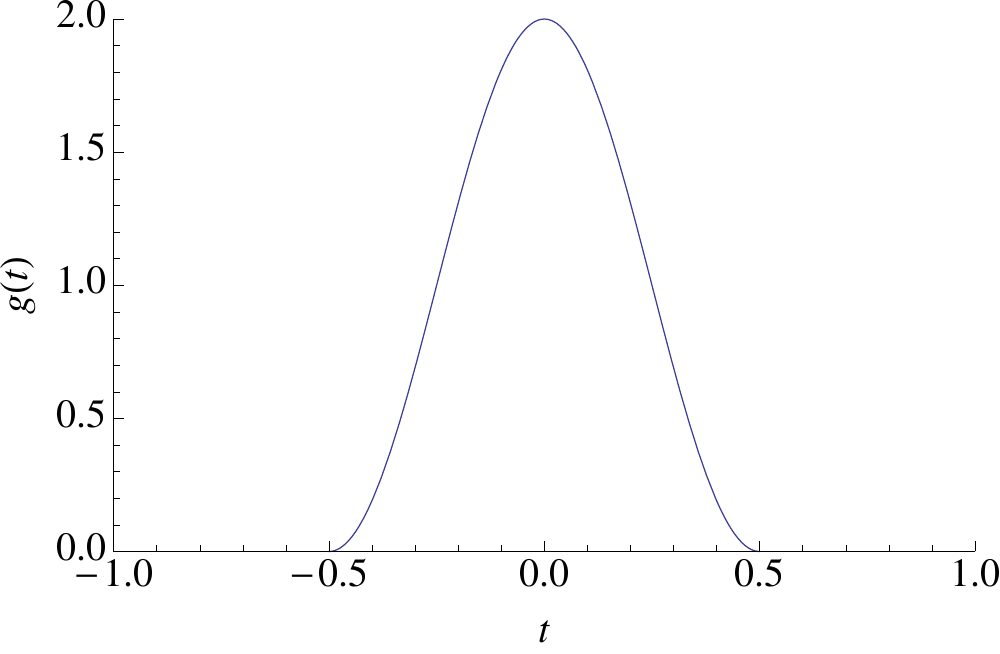}
\caption{\label{fig:g}(Color online) Normalized time-dependent system--apparatus coupling functions $g(t)$. The horizontal axis is in units of the measurement time $T$ and the vertical axis is in units of $1/T$. (a) Constant system--apparatus coupling $g(t)$ as defined in Eq.~\eqref{eq:igv34763578}. (b) Constant system--apparatus coupling with a linear turn-on and turnoff, each of duration $\Delta T$, shown here for $\Delta T/T=0.2$. (c) Triangular system--apparatus coupling, corresponding to a linear turn-on and turnoff, each of duration $\Delta T=T/2$. (d) System--apparatus coupling following a raised-cosine function as defined in Eq.~\eqref{eq:jfkhjkvhjkvhjkv11881}.}
\end{figure*}

First, we look at the case of constant $g(t)$ given by Eq.~\eqref{eq:igv34763578}, shown in Fig.~\ref{fig:g}(a). We may write $g(t)$ as $g(t) = \Pi(-T/2,T/2)$, where $\Pi(t_0,t_1)$ is the unit-area boxcar function defined by
\begin{align}\label{eq:igv}
\Pi(t_0,t_1) = \begin{cases} \frac{1}{t_1-t_0}, & t_0 \le t \le t_1 \quad (t_1>t_0) \\ 0 & \text{otherwise}. \end{cases}
\end{align}
We expect that the case of constant $g(t)$ is suboptimal in the sense that approximating the sharp corners of $g(t)$, which correspond to an infinitely fast turn-on and turnoff of the interaction, will require a broad spectrum of Fourier frequency components, leading to a large domain over which the Fourier transform $\widetilde{g}(\omega_{mn}T)$ of $g(t)$ exhibits a non-negligible amplitude. The Fourier transform $\widetilde{g}(\omega_{mn}T)$ [Eq.~\eqref{eq:lvdgs}] is readily evaluated, 
\begin{align}\label{eq:dklvdflk1}
\widetilde{g}(\omega_{mn}T) &= \frac{1}{T} \int_{-T/2}^{T/2} \D t\, \cos(\omega_{mn} t) = \mathrm{sinc}\left(\omega_{mn}T/2\right),
\end{align}
where $\mathrm{sinc}(x)=\sin(x)/x$ is the sinc function. Thus, the first-order transition amplitude is
\begin{align}\label{eq:dklvdflk2}
A^{(1)}_{m}(T) &= -\frac{\I}{\hbar} \bra{m} \op{O} \ket{n} \mathrm{sinc}\left(\omega_{mn}T/2\right),
\end{align}
and the corresponding  transition probability is
\begin{align}\label{eq:ubdv1}
\mathcal{P}^{(1)}_{m}(T) &= \frac{1}{\hbar^2}\abs{ \bra{m} \op{O} \ket{n}}^2 \mathrm{sinc}^2\left(\omega_{mn}T/2\right).
\end{align}

\begin{figure}
\begin{flushleft}
\hspace*{.2cm} (a)\vspace{-.4cm}
\end{flushleft}

\includegraphics[scale=.75]{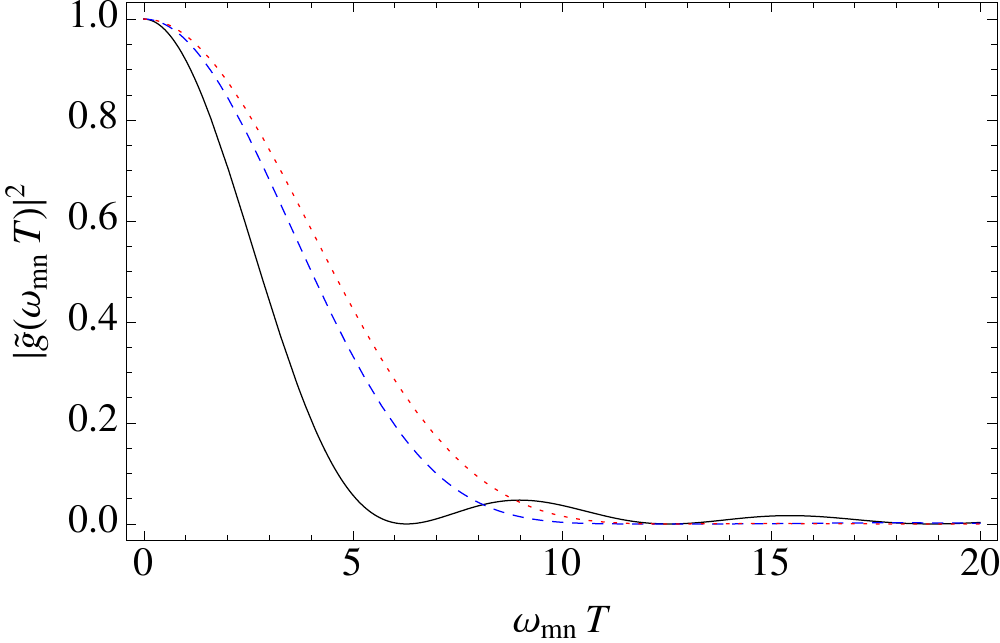}

\begin{flushleft}
\hspace*{.2cm} (b)\vspace{-.4cm}
\end{flushleft}

\includegraphics[scale=.75]{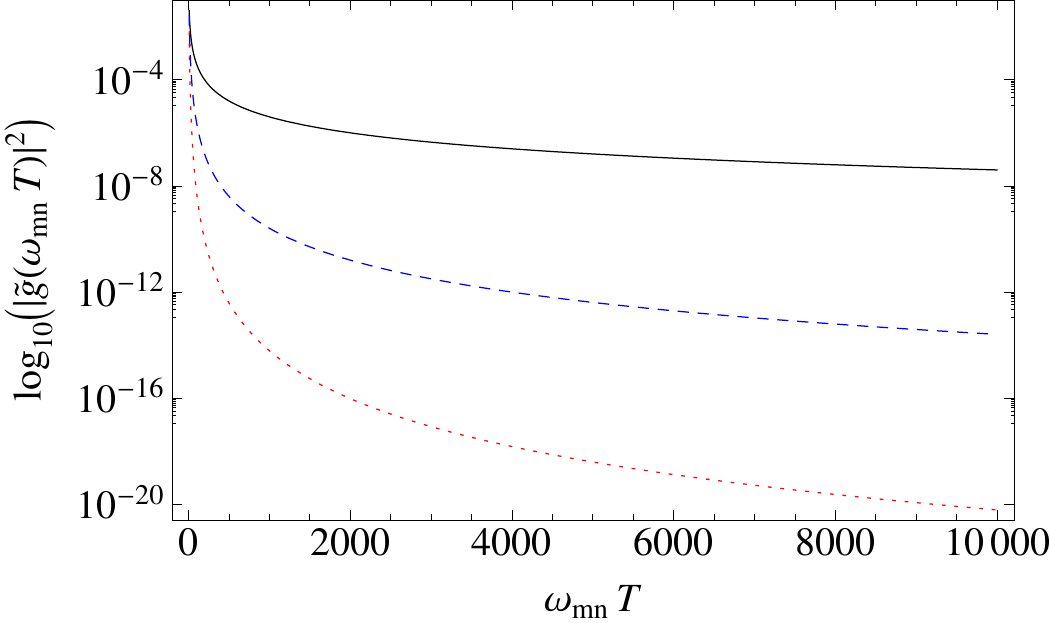}
\caption{\label{fig:ftcomp}(Color online) Comparison of the squared Fourier transform $\abs{\widetilde{g}(\omega_{mn}T)}^2$ for the following system--apparatus coupling functions $g(t)$ shown in Fig.~\ref{fig:g}: constant coupling $g(t)=1/T$ given by Eq.~\eqref{eq:igv34763578} (solid line), triangular $g(t)$ given by Eq.~\eqref{eq:178y45y} with $\Delta T=T/2$ (dashed line), and raised-cosine function given by Eq.~\eqref{eq:jfkhjkvhjkvhjkv11881} (dotted line). The dimensionless quantity $\abs{\widetilde{g}(\omega_{mn}T)}^2$ describes the dependence of the transition probability $\mathcal{P}^{(1)}_{m}(T)$ on the dimensionless parameter $\omega_{mn}T$, which quantifies the ratio of the measurement time $T$ to the time scale $\omega_{mn}^{-1}$ associated with the transition $\ket{n}\rightarrow\ket{m}$. (a) Plot of $\abs{\widetilde{g}(\omega_{mn}T)}^2$ for $T \gtrsim \omega_{mn}^{-1}$. (b) Decay of the envelope of $\abs{\widetilde{g}(\omega_{mn}T)}^2$ for $T \gg \omega_{mn}^{-1}$.}
\end{figure}

The dependence of $\abs{\widetilde{g}(\omega_{mn}T)}^2$, and therefore of $\mathcal{P}^{(1)}_{m}(T)$, on $\omega_{mn}T$ is shown in Fig.~\ref{fig:ftcomp} (solid line) separately for two regimes. Figure~\ref{fig:ftcomp}(a) shows the behavior for $T \gtrsim \omega_{mn}^{-1}$, while Fig.~\ref{fig:ftcomp}(b) shows the decay of the envelope of  $\abs{\widetilde{g}(\omega_{mn}T)}^2$ in the large-$T$ regime $T \gg \omega_{mn}^{-1}$ typically relevant to protective measurement. Assuming $T \gtrsim \omega_{mn}^{-1}$, the envelope of the function $\mathrm{sinc}^2 \left(\omega_{mn}T/2\right)$ decays as $(\omega_{mn}T/2)^{-2}$. Then, disregarding the oscillations of this function, we can approximate the transition-probability function \eqref{eq:ubdv1} by its envelope,
\begin{align}
\mathcal{P}^{(1)}_{m}(T) &\approx \frac{1}{\hbar^2}\abs{ \bra{m} \op{O} \ket{n}}^2 \frac{1}{(\omega_{mn}T/2)^2},
\end{align}
which is shown in Fig.~\ref{fig:ftcomp}(b). This demonstrates that in order to avoid appreciable state disturbance, i.e., $\mathcal{P}^{(1)}_{m}(T) \ll 1$, we must have
\begin{align}\label{eq:ubdv}
T \gg \frac{\abs{ \bra{m} \op{O} \ket{n}}}{\abs{E_m-E_n}} \qquad \text{for all $m\not= n$}.
\end{align}
For fixed (nonzero) values of the matrix elements $\bra{m} \op{O} \ket{n}$ and for an initial state equal to any energy eigenstate of the system, this condition simply means that $T$ must be significantly larger than the time scale set by the frequencies of the transitions between the different energy levels in the system. Not surprisingly, a condition of the form given in Eq.~\eqref{eq:ubdv} also follows from time-independent perturbation theory by imposing the requirement that the first-order state correction be small. 

We may also look at the width of the main peak (around $\omega_{mn}T=0$) of the function $\abs{\widetilde{g}(\omega_{mn} T)}^2=\mathrm{sinc}^2 \left(\omega_{mn}T/2\right)$. Note that in Fig.~\ref{fig:ftcomp}, $\abs{\widetilde{g}(\omega_{mn} T)}^2$ is plotted only for positive $\omega_{mn}T$ since $\abs{\widetilde{g}(\omega_{mn} T)}^2$ is even and thus only half of the main peak centered at $\omega_{mn} T=0$ is shown; we shall nonetheless refer to it as the central peak in the following. The function $\mathrm{sinc}^2(cx)$ has a full width at half maximum (FWHM) of $R(c) \simeq 2.78c^{-1}$ and thus the FWHM of the central peak of $\abs{\widetilde{g}(\omega_{mn} T)}^2$ is
\begin{align}\label{eq:ubdv1jhfdds33sf}
R\simeq 5.56.
\end{align}
This means that it suffices for $T$ to be equal to a few multiples of the transition time scale $\omega_{mn}^{-1}$ to reach the region outside the central peak of the transition probability $\mathcal{P}^{(1)}_{m}(T)$, as can also be seen from Fig.~\ref{fig:ftcomp}(a). 

Incidentally, the expression for $\mathcal{P}^{(1)}_{m}(T)$ [Eq.~\eqref{eq:ubdv1}] provides a physical illustration of why the assumption of a nondegenerate spectrum of $\op{H}_S$ is important to a proper protective measurement (see Ref.~\cite{Dass:1999:az} for an analysis of the issue of degeneracies in protective measurement). Namely, suppose there exists a state $\ket{m}\not=\ket{n}$ with energy $E_m=E_n$. Then $\omega_{mn}=0$ and
\begin{align}\label{eq:ubdv1jhf}
\mathcal{P}^{(1)}_{m}(T) &= \frac{1}{\hbar^2}\abs{ \bra{m} \op{O} \ket{n}}^2.
\end{align}
Assuming the matrix element $\bra{m} \op{O} \ket{n}$ does not vanish, this means that the probability for such an energy-conserving transition would have a nonzero value independent of $T$. In a protective measurement, however, the goal is to make this probability arbitrarily small by increasing $T$. The same observation also holds for the time-dependent couplings $g(t)$ studied below; in each case the Fourier transform of $g(t)$, and therefore the first-order transition probability $\mathcal{P}^{(1)}_{m}(T)$, becomes independent of $T$ if $\omega_{mn}=0$.

\subsection{\label{sec:turnonoff}Constant coupling with linear turn-on and turn-off}

Since the constant-coupling function $g(t)=1/T$ [Eq.~\eqref{eq:igv34763578}] exhibits a sharp step discontinuity at $\pm T/2$, we will now replace this discontinuity with a linear turn-on and turn-off of the measurement interaction over a period $\Delta T \le T/2$, and we will explore how this choice can help reduce the amount of state disturbance. Then $g(t)$ takes the shape of an isosceles trapezoid with base lengths $T$ and $T-2\Delta T$ [see Fig.~\ref{fig:g}(b)]. This function is equal to the convolution of two unit-area boxcar functions [see Eq.~\eqref{eq:igv}] of widths $\Delta T$ and $T-\Delta T$ centered at zero,
\begin{align}\label{eq:178y45y}
g(t) &=  \Pi\left(-\Delta T/2;\Delta T/2\right) \notag \\ &\quad * \, \Pi\left[-(T-\Delta T)/2;(T-\Delta T)/2\right].
\end{align}
Since the Fourier transform of the convolution of two functions is equal to the product of the Fourier transforms of each function, we have
\begin{align}\label{eq:hjvsjbfhj}
\widetilde{g}(\omega_{mn} T) &= \mathrm{sinc}\left(\omega_{mn}\Delta T/2\right)
\mathrm{sinc}\left[\omega_{mn}(T-\Delta T)/2\right],
\end{align}
and therefore the first-order transition probability is 
\begin{align}\label{eq:hjvsjbfhjdfh}
\mathcal{P}^{(1)}_{m}(T) &=\frac{1}{\hbar^2}\abs{ \bra{m} \op{O} \ket{n}}^2 \bigl\{\mathrm{sinc}\left(\omega_{mn}\Delta T/2\right)
\notag \\ &\quad\,\, \times \mathrm{sinc}\left[\omega_{mn}(T-\Delta T)/2\right]\bigr\}^2.
\end{align}
Increasing $\Delta T$ increases the rate of decay of the envelope of this function, but it also increases the FWHM. We shall explore two limiting cases. Setting $\Delta T =0$ corresponds to the case of constant $g(t)=1/T$ discussed in Sec.~\ref{sec:time-indep-coupl} and Eq.~\eqref{eq:hjvsjbfhj} becomes $\widetilde{g}(\omega_{mn} T) = 
\mathrm{sinc} \left(\omega_{mn}T/2\right)$, in agreement with Eq.~\eqref{eq:dklvdflk1}. The other limiting case corresponds to setting $\Delta T =T/2$. Then $g(t)$ becomes the unit-area triangle function shown in Fig.~\ref{fig:g}(c) and Eq.~\eqref{eq:hjvsjbfhj} gives
\begin{align}\label{eq:hjvsjbfhjdkhsf2}
\widetilde{g}(\omega_{mn} T) &= \mathrm{sinc}^2\left(\omega_{mn}T/4\right).
\end{align}
Note that this function reaches a maximum value of $2/T$ at $t=0$, which is twice the value for constant $g(t)$ [see Eq.~\eqref{eq:igv34763578}]. The corresponding first-order transition probability is 
\begin{align}\label{eq:hjvsjb44fh}
\mathcal{P}^{(1)}_{m}(T) = \frac{1}{\hbar^2}\abs{ \bra{m} \op{O} \ket{n}}^2 \mathrm{sinc}^4\left(\omega_{mn}T/4\right).
\end{align}
The dependence of this function on $\omega_{mn}T$ is shown in Fig.~\ref{fig:ftcomp} (dashed line). The envelope decays as $1/(\omega_{mn}T/4)^4$, two orders faster than in the case of constant $g(t)$. This is a significant improvement in the decay rate. The longer we make the turn-on and turnoff periods $\Delta T$, i.e., the more we approach the limiting case of the triangle function $g(t)$ shown in Fig.~\ref{fig:g}(c), the more quickly the envelope of the transition probability decreases with $T$. Note that the condition on the relationship between $T$ and $\omega_{mn}^{-1}$ stated in Eq.~\eqref{eq:ubdv} still applies here; the only difference is that $\abs{ \bra{m} \op{O} \ket{n}}$ in Eq.~\eqref{eq:ubdv} is replaced by $\abs{ \bra{m} \op{O} \ket{n}}^{1/2}$. 

Finally, since the FWHM of the function $\mathrm{sinc}^4(cx)$ is $R(c) \simeq 2.00c^{-1}$, the FWHM of the central peak of $\abs{\widetilde{g}(\omega_{mn} T)}^2$ is [see also  Fig.~\ref{fig:ftcomp}(a)]
\begin{align}\label{eq:643tfguwd}
R\simeq 8.00.
\end{align}
Given that in a protective measurement we typically have $\omega_{mn}T \gg 1$ and are therefore far away from the central peak, the increase in the FWHM compared to the case of constant $g(t)$ (which was $R\simeq 5.56$) can be considered irrelevant.

\subsection{\label{sec:smoothly-vary-coupl}Smoothly varying coupling} 

We saw in Sec.~\ref{sec:turnonoff} that inclusion of linear turn-on and turnoff periods significantly decreases the state disturbance. Even a triangular $g(t)$, however, has sharp corners at the turn-on and turnoff points $t=\pm T/2$ as well as at the center $t=0$; these corners can be expected to contribute additional Fourier components and therefore increase the bandwidth. In the following we shall therefore consider a smoothly varying coupling function $g(t)$ without sharp corners. We choose the raised-cosine function with unit area shown in Fig.~\ref{fig:g}(d),
\begin{equation}\label{eq:jfkhjkvhjkvhjkv11881}
g(t)=\begin{cases} \frac{1}{T}\left[ 1+\cos(2\pi t/T)\right], & -T/2 \le t \le T/2 \\ 0 & \text{otherwise}. \end{cases}
\end{equation}
As in the case of triangular $g(t)$, this function has a maximum value of $2/T$ at $t=0$. The Fourier transform is
\begin{equation}\label{eq:12786svjj}
\widetilde{g}(\omega_{mn}T)=\frac{1}{1-(\omega_{mn} T/ 2\pi)^2} \mathrm{sinc}(\omega_{mn} T/2).
\end{equation}
Comparing this result to the Fourier transform of constant $g(t)=1/T$ [see Eq.~\eqref{eq:dklvdflk1}], we see that we have gained an extra factor of $\left[1-(\omega_{mn} T/ 2\pi)^2\right]^{-1}$. The first-order transition probability is
\begin{align}
\mathcal{P}^{(1)}_m(T) &= \frac{1}{\hbar^2} \abs{ \bra{m} \op{O} \ket{n}}^2 \notag \\ &\quad \times \left[\frac{1}{1-(\omega_{mn} T/ 2\pi)^2} \mathrm{sinc}(\omega_{mn} T/2) \right]^2.
\end{align}
Its dependence on $\omega_{mn}T$ is shown in Fig.~\ref{fig:ftcomp} (dotted line). Note that it decays as $1/T^6$ for large $\omega_{mn} T$. This is to be compared to the $1/T^2$ dependence for constant $g(t)$ [see Eq.~\eqref{eq:ubdv1}] and the $1/T^4$ dependence for triangular $g(t)$ [see Eq.~\eqref{eq:hjvsjb44fh}]. Thus, choosing a smoothly changing $g(t)$ provides a decisive advantage in reducing the state disturbance, in agreement with what one would generally expect in light of the quantum adiabatic theorem.

Since the FWHM of the function
\begin{equation}
f(x)=\left[\frac{1}{1-(cx/\pi)^2}\mathrm{sinc}(cx)\right]^2
\end{equation}
is $R(c) \simeq 4.53c^{-1}$, the FWHM of $\abs{\widetilde{g}(\omega_{mn}T)}^2$, with $\widetilde{g}(\omega_{mn}T)$ given by Eq.~\eqref{eq:12786svjj}, is
\begin{equation}
R \simeq 9.06.
\end{equation}
Table~\ref{tab:comp} summarizes the results for the scaling behavior and FWHM of the transition probability $\mathcal{P}^{(1)}_m(T)$ for three different choices of $g(t)$. Note the dramatic difference in the falloff of the probability with $T$. By contrast, the FWHM is of order unity in all cases and the differences in the specific values are insignificant. 

\begin{table}
\begin{ruledtabular}
\begin{tabular}{ccc}
& Transition probability & FWHM of $\mathcal{P}^{(1)}_m(T)$ \\
Coupling $g(t)$ & $\mathcal{P}^{(1)}_m(T)$ & (in units of $\omega_{mn}T$) \\\hline 
constant & $O(1/T^2)$ & $5.56$ \\
triangle & $O(1/T^4)$ & $8.00$ \\
raised cosine & $O(1/T^6)$ & $9.06$ \\
\end{tabular}
\end{ruledtabular}
\caption{\label{tab:comp}Comparison of the dependence of the transition probability $\mathcal{P}^{(1)}_m(T)$ and its peak width (FWHM) on the measurement time $T$ for three different choices of the coupling function $g(t)$.}
\end{table}

\subsection{Higher-order contributions}

To complete our analysis, we may also look at the higher-order contributions $A^{(\ell \ge 2)}_{m}(T)$ to the transition amplitude $A_{m}(T)$, which are given by Eq.~\eqref{eq:g8fbvsv1}. As shown in Appendix~\ref{sec:high-order-contr}, while such higher-order contributions turn out to contain terms that are of the same order in $1/T$ and show similar $T$ dependence as the first-order amplitude $A^{(1)}_{m}(T)$, these terms becomes exponentially suppressed with increasing order. Therefore, the first-order amplitude $A^{(1)}_{m}(T)$ discussed here provides a good representation and approximation of the full transition amplitude and the state disturbance incurred in a protective measurement.

\section{\label{sec:pointer-shift}Pointer shift}

In Sec.~\ref{sec:state-disturbance} we showed how choosing a smoothly varying coupling function $g(t)$ can significantly improve the rate at which the state disturbance decreases as the measurement time $T$ is increased. Based on this result alone, we would conclude that the optimal choice of $g(t)$ is a functional form that minimizes the state disturbance in this sense. However, there is another concern to be taken into account in protective measurement, namely, the shift of the apparatus pointer. Specifically, the question is whether and how the particular choice of $g(t)$ may influence the amount by which the apparatus pointer will move during the measurement time $T$. 

We will now show that in the relevant large-$T$ case given by Eq.~\eqref{eq:ubdv}, the pointer shift is independent of the particular form of $g(t)$. To this end, we quantify the pointer shift associated with the initial state $\ket{n}$ of the system by perturbatively studying the amplitude of this state at the conclusion of the protective measurement. As before, we let the interaction start at $t=-T/2$ and end at $t=T/2$. From Eq.~\eqref{eq:87gr782} the first-order contribution to the amplitude is 
\begin{equation}
A^{(1)}_{n}(T) = -\frac{\I}{\hbar} \bra{n} \op{O} \ket{n} \int_{-T/2}^{T/2} \D t\, g(t)  = -\frac{\I}{\hbar} \bra{n} \op{O} \ket{n},
\end{equation}
where the last step follows from the normalization of $g(t)$ according to Eq.~\eqref{eq:normal}. Thus, $A^{(1)}_{n}(T)$ is independent of the particular functional form of $g(t)$. We will now investigate the higher-order terms $A^{(\ell \ge 2)}_{n}(T)$ given by Eq.~\eqref{eq:g8fbvsv1} with $m=n$,
\begin{align}\label{eq:jldfjkl2}
A^{(\ell)}_{n}(T) &= 
\left(-\frac{\I}{\hbar}\right)^\ell \sum_{k_1,k_2,\hdots,k_{\ell-1}}
\bra{n} \op{O} \ket{k_1} \bra{k_1} \op{O} \ket{k_2} \notag \\ &\quad \times \bra{k_2} \op{O} \ket{k_3} \cdots \bra{k_{\ell-1}} \op{O} \ket{n}\notag \\ &\quad \times\int_{-T/2}^{T/2} \D t' \, \E^{\I \omega_{nk_1} t'} g(t') \notag \\ &\quad \times\int_{-T/2}^{t'} \D t'' \,\E^{\I \omega_{k_1k_2} t''} g(t'')\cdots \notag \\ &\quad \times\int_{-T/2}^{t^{(\ell-1)}} \D t^{(\ell)} \,\E^{\I \omega_{k_{\ell-1} n} t^{(\ell)}}g(t^{(\ell)}).
\end{align}
Let us consider the term for which all $k_j=n$, $1 \le j \le \ell-1$, 
\begin{align}\label{eq:1aa}
A^{(\ell)}_{n,\{k_j=n\}}(T) &= \left(-\frac{\I}{\hbar}\right)^\ell 
\bra{n} \op{O} \ket{n}^\ell \int_{-T/2}^{T/2} \D t' \, g(t') \notag \\ &\quad \times\int_{-T/2}^{t'} \D t'' \, g(t'') \cdots \int_{-T/2}^{t^{(\ell-1)}} \D t^{(\ell)} \, g(t^{(\ell)}).
\end{align}
In Appendix~\ref{sec:eval-point-shift} we show that the multiple integral is equal to $1/\ell !$ and thus Eq.~\eqref{eq:1aa} becomes
\begin{align}\label{eq:dsfdf1aa}
A^{(\ell)}_{n,\{k_j=n\}}(T) &= \frac{1}{\ell !}\left(-\frac{\I}{\hbar}\right)^\ell 
\bra{n} \op{O} \ket{n}^\ell.
\end{align}
To find the total amplitude that also includes the effect on the apparatus subspace, we use Eq.~\eqref{eq:y89ggy892} and sum $A^{(\ell)}_{n,\{k_j=n\}}(T)$ over all orders $\ell$,
\begin{align}\label{eq:vcsuaadfv}
\sum_{\ell=0}^\infty a_i^{\ell}  A^{(\ell)}_{n,\{k_j=n\}}(T) &= \sum_{\ell=0}^\infty \frac{1}{\ell !} \left( -\frac{\I}{\hbar} a_i \bra{n} \op{O} \ket{n} \right)^\ell \notag \\ &=
\E^{-\I a_i\bra{n} \op{O} \ket{n}/\hbar} \notag \\ &= \bra{A_i} \E^{-\I \op{P}\bra{n} \op{O} \ket{n}/\hbar} \ket{A_i},
\end{align}
where $\E^{-\I \op{P}\bra{n} \op{O} \ket{n}/\hbar}$ is the familiar result for the pointer-shift operator in zeroth-order protective measurement obtained from time-independent perturbation theory [see Eq.~\eqref{eq:isvgaazxxx}]. Note that we must include all orders $\ell$ irrespective of the size of $T$ because each of the $\ell$ matrix elements $\bra{k_j} g(t)\op{O} \ket{k_{j'}}$ is of order $1/T$ and the integrals combined are of order $T^\ell$. 

Note that if $g(t)$ is not normalized, i.e., if $G\equiv\int_{-T/2}^{T/2} \D t \, g(t) \not= 1$, then the multiple integral in Eq.~\eqref{eq:1aa} is equal to $G^\ell/\ell !$ and Eq.~\eqref{eq:vcsuaadfv} instead reads
\begin{align}\label{eq:dsfdf1aaadsdd}
\sum_{\ell=0}^\infty a_i^\ell A^{(\ell)}_{n,\{k_i=n\}}(T) &=\E^{-\I G a_i\bra{n} \op{O} \ket{n}/\hbar} \notag \\ &= \bra{A_i}\E^{-\I G\op{P}\bra{n} \op{O} \ket{n}/\hbar}\ket{A_i}.
\end{align}
Thus, the pointer shift is proportional to both $\bra{n} \op{O} \ket{n}$ and $G$, the area under the $g(t)$ graph. 

As shown in Appendix~\ref{sec:corr-constant-coupling}, for constant $g(t)$ the $k_j\not=n$ contributions to Eq.~\eqref{eq:jldfjkl2} are of order $1/T$ or higher. As far as the problem of the pointer shift is concerned, they may therefore be neglected in the large-$T$ limit of protective measurement given by Eq.~\eqref{eq:ubdv}. Choosing time-varying coupling functions $g(t)$ can only further diminish the relevance of these terms since we know from our analysis of state disturbance in Sec.~\ref{sec:state-disturbance} that such couplings will significantly increase the rate of amplitude decay with $T$.

\section{\label{sec:discussion}Discussion and conclusions}

This paper provides a quantitative and detailed analysis of the state disturbance incurred during a protective measurement under physically meaningful conditions, supplying knowledge that is crucial not only to a practical implementation of protective measurement, but also to a deeper understanding of the theory and dynamics of protective measurement. We have departed from the prevailing mathematical idealization of infinitely weak or perfectly adiabatic protective measurements and instead investigated the amount of state disturbance introduced by protective measurements characterized by a finite time-dependent system--apparatus coupling strength $g(t)$ and finite duration $T$. In studying the state disturbance, we have focused on the first-order transition probabilities obtained from time-dependent perturbation theory, which quantify the mixing of states of the system different from the initial state and are proportional to the squared Fourier transform of $g(t)$. For the functions $g(t)$ studied here, their Fourier transforms, and thus the corresponding transition probabilities, are functions of the dimensionless quantity $\omega_{mn}T$, which measures the ratio of the measurement time $T$ to the internal time scale $\omega_{mn}^{-1}$ associated with the transition $\ket{n}\rightarrow\ket{m}$, with $\ket{n}$ denoting the initial state of the system. 

The choice of constant $g(t)=1/T$ [with $g(t)=0$ outside the measurement interval] commonly considered in the literature (see, e.g., Refs.~\cite{Dass:1999:az,Vaidman:2009:po,Gao:2013:om}) is found to be an essentially worst-case scenario for state disturbance, owing to an infinitely fast turn-on and turnoff of the interaction that leads to a broad Fourier spectrum. In agreement with what one would expect from first-order time-independent perturbation theory, the envelope of the corresponding transition probabilities decays as $1/T^2$. The condition for these probabilities to be small is that the measurement $T$ must be (ideally significantly) larger than the longest internal time scale $\tau_\text{max} = \max_{m\not=n}\{\omega_{mn}^{-1}\}$ associated with perturbation-induced transitions out of the initial state $\ket{n}$. Given that typical atomic time scales are very short, this means that it need not be difficult in practice to choose the measurement time $T$ long enough to ensure sufficiently small state disturbance (see also Ref.~\cite{Dickson:1995:lm} for a similar argument).

We have shown that any smoothing of the coupling function $g(t)$ dramatically improves the rate of envelope decay of the transition probabilities with $T$ [see Fig.~\ref{fig:ftcomp}(b)]. For example, introduction of linear turn-on and turnoff periods increases the envelope decay rate to $1/T^\beta$, with $2 < \beta \le 4$. The longer we make the turn-on and turn-off periods, the more quickly the envelope of the transition probability decreases with $T$. The value $\beta=4$ is reached if the lengths of the turn-on and turn-off periods are maximized to $T/2$, i.e., if the interaction strength is linearly ramped up to its maximum value of $2/T$ and then immediately linearly decreased back down to zero, corresponding to a triangular pulse. Smoothing of $g(t)$ by using a raised-cosine function for $g(t)$ provides a further significant improvement, leading to a $1/T^6$ envelope decay of the transition probabilities. 

For all the choices of $g(t)$ considered in this paper, the width of the central dominant peak of the transition probability for $\ket{n}\rightarrow\ket{m}$, considered as a function of $\omega_{mn}T$, has a similar value lying between about 5 and 10. It is therefore sufficient for $T$ to be equal to just a few multiples of the internal time scales $\omega_{mn}^{-1}$ to reach a region outside the central peak where the transition probabilities become relatively small. The similarity of these values for the peak width suggests that we should let the choice of $g(t)$ chiefly be guided by the goal of maximizing the decay rate of the transition-probability envelope, which means choosing a smoothly varying $g(t)$ such as the raised-cosine function considered here. In the limit $T \rightarrow \infty$, both the raised-cosine function and triangular $g(t)$ change infinitely slowly and the transition probabilities are zero, providing a concrete illustration of the quantum adiabatic theorem \cite{Born:1928:yf}.

Interestingly, we found that higher-order corrections to the transition amplitude contain terms that exhibit the same scaling behavior with $T$ as the first-order contribution (e.g., proportional to $1/T$ for constant coupling). We attributed this observation to the particular way in which the duration $T$ of the protective-measurement interaction tunes the strength of the interaction via the normalization condition $\int \D t\, g(t)=1$. Such terms, however, become exponentially suppressed as one moves to higher-order corrections, suggesting that the first-order transition amplitude is indeed the dominant and appropriate quantity for measuring the state disturbance. 

We also showed that the total pointer shift incurred during the protective measurement is independent of the functional form of $g(t)$. To be sure, there are higher-order corrections to the evolution of the wave packet of the apparatus pointer (corresponding to effects such as additional spreading and distortion of the wave packet) whose precise magnitudes and dynamics will depend on the choice of $g(t)$. However, such corrections become insignificant in the case $T \gg \tau_\text{max}$ relevant to protective measurement. Moreover, their influence can be expected to be minimized by using a coupling function $g(t)$ that minimizes the state disturbance. Choosing a smoothly varying coupling function, such as the triangle function or the raised-cosine function discussed in this paper, has therefore two benefits: It reduces the disturbance of the initial state and it improves the approximation of treating the evolution of the pointer wave packet as a simple combination of free spreading and a shift of the center of the wave packet by an amount given by the expectation value of the measured observable $\op{O}$ in the initial state of the system.

While theoretical schemes for state reconstruction using protective measurements have been described for specific systems and system--apparatus interactions \cite{Aharonov:1993:jm,Anandan:1993:uu,Nussinov:1998:yy,Dass:1999:az}, the experimental realization of protective measurements remains an open challenge. We hope that our analysis of state disturbance, as well as our framework for treating arbitrary time-dependent coupling functions, may aid in the implementation of protective measurements. Since the system--apparatus interaction in weak measurements \cite{Aharonov:1988:mz} is of the same structure as in protective measurement, only of much shorter duration, our results may also be of interest to the theory and implementation of weak measurements.

\begin{acknowledgments} 
This research was supported by a University of Portland Arthur Butine Grant.
\end{acknowledgments} 

\appendix

\section{\label{sec:high-order-corr}Higher-order corrections to the state-vector amplitude}

Here we will investigate the contributions of the higher-order corrections $A^{(\ell \ge 2)}_{m}(T)$ [see Eq.~\eqref{eq:g8fbvsv1}] to the amplitude $A_{m}(T)$ and discuss their influence on the state disturbance (Appendix~\ref{sec:high-order-contr}) and the evolution of the pointer wave packet (Appendix~\ref{sec:corr-constant-coupling}).

\subsection{\label{sec:high-order-contr}State disturbance}

The higher-order contributions $A^{(\ell)}_{m}(T)$, $\ell \ge 2$, to the transition amplitude $A_{m}(T)$ are given by Eq.~\eqref{eq:g8fbvsv1} with $m\not= n$,
\begin{align}\label{eq:g8f4656bvsv1dfhjjhfssd}
A^{(\ell)}_{m}(T) &= \left(-\frac{\I}{\hbar}\right)^\ell \sum_{k_1,k_2,\hdots,k_{\ell-1}}
\bra{m} \op{O} \ket{k_1} \bra{k_1} \op{O} \ket{k_2} \cdots \notag \\ & \quad \times \bra{k_{\ell-1}} \op{O} \ket{n} \int_{-T/2}^{T/2}  \D t' \, \E^{\I \omega_{mk_1} t'} g(t') \notag \\ & \quad \times\int_{-T/2}^{t'} \D t'' \,\E^{\I \omega_{k_1k_2} t''} g(t'') \cdots \notag \\ & \quad \times\int_{-T/2}^{t^{(\ell-1)}} \D t^{(\ell)} \,\E^{\I \omega_{k_{\ell-1} n} t^{(\ell)}} g(t^{(\ell)}).
\end{align}
This amplitude represents multistep transitions in which the system transitions from the initial state $\ket{n}$ to the final state $\ket{m}$ via up to $\ell-1$ intermediate virtual states $\ket{k_i}$, $1 \le i \le \ell-1$, summed over all possible transition times and intermediate states.

We first consider the terms in the sum in Eq.~\eqref{eq:g8f4656bvsv1dfhjjhfssd} for which all indices $k_i$ take on distinct values; let us denote such a term by $\alpha^{(\ell)}_{k_1\cdots k_{\ell-1}}(T)$. (As always, we also assume that all transition frequencies $\omega_{k_ik_j}$ are nonzero for $k_i\not= k_j$, i.e., that the spectrum is nondegenerate.) To estimate the influence of these terms, we approximate the frequencies $\omega_{k_ik_j}$ by a typical value $\bar{\omega}$ such that we can write $\alpha^{(\ell)}_{k_1\cdots k_{\ell-1}}(T)$ as
\begin{align}\label{eq:g8fbvsv1dfhjjhfd}
\alpha^{(\ell)}_{k_1\cdots k_{\ell-1}}(T) &= \left(-\frac{\I}{\hbar}\right)^\ell 
\bra{m} \op{O} \ket{k_1} \bra{k_1} \op{O} \ket{k_2} \cdots \bra{k_{\ell-1}} \op{O} \ket{n}\notag \\ &\quad\times \int_{-T/2}^{T/2}  \D t' \, \E^{\I \bar{\omega} t'} g(t') \notag \\ &\quad\times\int_{-T/2}^{t'} \D t'' \,\E^{\I \bar{\omega} t''} g(t'') \cdots \notag \\ &\quad\times  \int_{-T/2}^{t^{(\ell-1)}} \D t^{(\ell)} \,\E^{\I \bar{\omega} t^{(\ell)}} g(t^{(\ell)}).
\end{align}
Since the integrands are symmetric under an exchange of the time variables, we can employ the strategy described in Appendix~\ref{sec:eval-point-shift} below. Namely, we can replace the upper integration limits by $T/2$ and compensate for this modification by an overall multiplicative factor of $1/\ell!$,
\begin{align}\label{eq:g8fbvsv1dfhjjhfssd}
\alpha^{(\ell)}_{k_1\cdots k_{\ell-1}}(T) &= \left(-\frac{\I}{\hbar}\right)^\ell 
\bra{m} \op{O} \ket{k_1} \bra{k_1} \op{O} \ket{k_2}\cdots \bra{k_{\ell-1}} \op{O} \ket{n} \notag \\ &\quad\times \frac{1}{\ell!} \left[ \int_{-T/2}^{T/2}  \D t' \, \E^{\I \bar{\omega} t'} g(t') \right]^\ell.
\end{align}
The term in square brackets is simply the Fourier transform $\widetilde{g}(\bar{\omega};T)$ of $g(t)$ [compare Eq.~\eqref{eq:lvdgs}]. Thus, our analysis of the first-order contributions in Secs.~\ref{sec:time-indep-coupl}--\ref{sec:smoothly-vary-coupl} can be directly applied to determining the dependence of Eq.~\eqref{eq:g8fbvsv1dfhjjhfssd} on the measurement time $T$. Specifically, if the first-order transition amplitude $A^{(1)}_{m}(T)$ follows a $1/T^\beta$ dependence, then the contribution to the transition amplitude $A^{(\ell)}_{m}(T)$ made by Eq.~\eqref{eq:g8fbvsv1dfhjjhfssd} will be on the order of $1/T^{\beta \ell}$. Since $\ell \ge 2$, such contributions are negligible in the large-$T$ limit relevant to protective measurement. 

It follows that contributions whose dependence on $T$ is comparable to that of the first-order contribution $A^{(1)}_{m}(T)$ can only arise, if at all, for transitions involving less than $\ell-1$ intermediate virtual states $\ket{k\not=m,n}$. To this end, let us consider the terms in which no such distinct virtual intermediate transitions are present. Formally, this corresponds to setting $k_j=m$ for $1 \le j \le i$ and $k_j=n$ for $i+1 \le j \le \ell-1$ in Eq.~\eqref{eq:g8f4656bvsv1dfhjjhfssd}, with $1 \le i \le \ell-1$. There are $\ell$ such terms, which all give rise to the same dependence on $T$. Let us examine a representative term, 
\begin{align}\label{eq:ub22dv2hh437tyr67}
\alpha^{(\ell)}_{mn\cdots n}(T) &= \left(-\frac{\I}{\hbar}\right)^\ell \bra{m} \op{O} \ket{n} \bra{n} \op{O} \ket{n}^{\ell-1} \notag \\ & \quad\times \int_{-T/2}^{T/2}  \D t' \, \E^{\I \omega_{mn} t'} g(t') \int_{-T/2}^{t'} \D t'' \,g(t'') \cdots \notag \\ &\quad\times \int_{-T/2}^{t^{(\ell-1)}} \D t^{(\ell)} \, g(t^{(\ell)}).
\end{align}
Again using the strategy described in Appendix~\ref{sec:eval-point-shift}, we rewrite this equation as
\begin{align}\label{eq:ubh437tyr6aaww7aa}
\alpha^{(\ell)}_{mn\cdots n}(T) &= \left(-\frac{\I}{\hbar}\right)^\ell \bra{m} \op{O} \ket{n} \bra{n} \op{O} \ket{n}^{\ell-1}\notag \\ & \quad \times \int_{-T/2}^{T/2}  \D t' \, \E^{\I \omega_{mn} t'} g(t') \frac{1}{(\ell-1)!}\left[G(t')\right]^{\ell-1},
\end{align}
where $G(t) = \int_{-T/2}^{t} \D t' \,g(t')$ measures the area under the $g(t)$ curve in the interval $[-T/2,t']$, with $-T/2 \le t' \le T/2$; this area is a non-negative dimensionless number of order unity. Thus, $\alpha^{(\ell)}_{mn\cdots n}(T)$ is of the same order in $1/T$ as the integral $\int_{-T/2}^{T/2}  \D t' \, \E^{\I \omega_{mn} t'} g(t')$, which is the Fourier transform of $g(t)$. We know from Secs.~\ref{sec:time-indep-coupl}--\ref{sec:smoothly-vary-coupl} that this Fourier transform determines the $T$ dependence of the first-order transition amplitude $A^{(1)}_{m}(T)$. Therefore, we can conclude that $\alpha^{(\ell)}_{mn\cdots n}(T)$ [Eq.~\eqref{eq:ub22dv2hh437tyr67}] has a similar $T$ dependence as $A^{(1)}_{m}(T)$. Specifically, just like $A^{(1)}_{m}(T)$, it scales as $1/T$ for constant coupling (Sec.~\ref{sec:time-indep-coupl}), $1/T^2$ for constant coupling with linear turn-on and turnoff (Sec.~\ref{sec:turnonoff}), and $1/T^3$ for the raised-cosine coupling function (Sec.~\ref{sec:smoothly-vary-coupl}). 

Instead of simply estimating the influence of $G(t)$ in Eq.~\eqref{eq:ubh437tyr6aaww7aa} by unity, as we have just done, let us also explicitly evaluate this function for the case of constant $g(t)$. To simplify notation, let us take $g(t)=1/T$ in the interval $[0,T]$ rather than $[-T/2,T/2]$. Then $G(t)=t/T$ and Eq.~\eqref{eq:ubh437tyr6aaww7aa} becomes
\begin{align}\label{eq:ubh437tyh7r67aaxxx}
\alpha^{(\ell)}_{mn\cdots n}(T) &= \left(-\frac{\I}{\hbar}\right)^\ell \bra{m} \op{O} \ket{n} \bra{n} \op{O} \ket{n}^{\ell-1} \frac{1}{(\ell-1)!} \frac{1}{T^{\ell-1}} \notag \\ &\times \int_{0}^{T}  \D t \, \E^{\I \omega_{mn} t} g(t) t^{\ell-1}.
\end{align}
The integral is the Fourier transform of the function $g(t) t^{\ell-1}$, which is given by
\begin{equation}
\int_{0}^{T}  \D t \, \E^{\I \omega_{mn} t} g(t) t^{\ell-1} = \I^{\ell-1} \frac{\D^{\ell-1} \widetilde{g}(\omega_{mn};T)}{\D \omega_{mn}^{\ell-1}},
\end{equation}
where $\widetilde{g}(\omega_{mn};T)=\int_{0}^{T}  \D t \, \E^{\I \omega_{mn} t} g(t)$ is the Fourier transform of $g(t)$. By the shift property of the Fourier transform, $\widetilde{g}(\omega_{mn};T)$ is equal to the Fourier transform of the original $g(t)$ defined on the interval $[-T/2,T/2]$ save for an overall phase factor $\E^{-\I \omega_{mn}T/2}$ [this result holds for arbitrary functions $g(t)$, including the ones considered in Secs.~\ref{sec:turnonoff} and \ref{sec:smoothly-vary-coupl}]. Thus, the $T$ dependence of $\alpha^{(\ell)}_{mn\cdots n}(T)$ [Eq.~\eqref{eq:ubh437tyh7r67aaxxx}] is given by
\begin{equation}\label{eq:ubh437tyh7r67aax666}
\gamma_\ell(\omega_{mn};T)=\frac{1}{T^{\ell-1}} \frac{\D^{\ell-1} \widetilde{g}(\omega_{mn};T)}{\D \omega_{mn}^{\ell-1}},
\end{equation}
where we can now take $\widetilde{g}(\omega_{mn};T)$ to denote the Fourier transform of $g(t)$ defined on the original interval $[-T/2,T/2]$.
For constant $g(t)$, $\widetilde{g}(\omega_{mn};T)\equiv \widetilde{g}(\omega_{mn}T) = \mathrm{sinc}\left(\omega_{mn}T/2\right)$ [see Eq.~\eqref{eq:dklvdflk1}]. In the expression for $\frac{\D^n}{\D x^n} \mathrm{sinc}(ax)$, the term to leading order in $a$ is equal to $\pm a^n \mathrm{sinc}(ax)$ if $n$ is even and equal to $\pm a^n \frac{\cos(ax)}{ax}$ if $n$ is odd. Therefore, to leading order in $1/T$, Eq.~\eqref{eq:ubh437tyh7r67aax666} can be approximated by
\begin{equation}\label{eq:ubh437tyx666}
\gamma_\ell(\omega_{mn};T)= \begin{cases} \pm \mathrm{sinc}\left(\omega_{mn}T/2\right), & \ell=3,5,7,\hdots \\
\pm\frac{\cos \left(\omega_{mn}T/2\right)}{\omega_{mn}T/2}, & \ell=2,4,6,\hdots\,. \end{cases}
\end{equation}
When $T$ is significantly larger than $\omega_{mn}$ (which is the case relevant to protective measurement), both $\mathrm{sinc}\left(\omega_{mn}T/2\right)$ and $\frac{\cos \left(\omega_{mn}T/2\right)}{\omega_{mn}T/2}$ exhibit similar behavior; in particular, their envelopes decay as $1/T$. We have thus confirmed our previous result that the term $\alpha^{(\ell)}_{mn\cdots n}(T)$, Eq.~\eqref{eq:ub22dv2hh437tyr67}, exhibits essentially the same $T$-dependence as the first-order amplitude $A^{(1)}_{m}(T)$.

Note, however, that the factor $1/(\ell-1)!$ appearing in Eqs.~\eqref{eq:ubh437tyr6aaww7aa} and \eqref{eq:ubh437tyh7r67aaxxx} exponentially damps such first-order-like terms with increasing order $\ell$ (while their number grows only linearly with $\ell$). This implies that their contributions rapidly become insignificant with increasing $\ell$. Also, since we have shown that their $T$ dependence is well approximated by the Fourier transform of $g(t)$, which determines the $T$ dependence of the first-order amplitude $A^{(1)}_{m}(T)$, it follows that, to leading order in $1/T$, $A^{(1)}_{m}(T)$ is indeed a good representation of the dependence of the overall transition probability on $T$. This result justifies our focus on $A^{(1)}_{m}(T)$ in Secs.~\ref{sec:time-indep-coupl}--\ref{sec:smoothly-vary-coupl}.

So far we have considered the two limiting cases of a maximum number of virtual transitions $\ket{k\not=m,n}$ and no virtual intermediate transitions at all. Applying our above reasoning to terms corresponding to the intermediate regime in which there is a nonzero but nonmaximum number of virtual transitions, it is readily seen that such terms must at least be of order $1/T^{2\beta}$ or higher, where $\beta$ is the scaling parameter for the envelope decay of the first-order amplitude $A^{(1)}_{m}(T)$. Therefore, these terms can be neglected compared to the $O(1/T^\beta)$ terms arising from  $A^{(1)}_{m}(T)$ and from the first-order-like contributions to the higher-order amplitudes $A^{(\ell)}_{m}(T)$ discussed above.

As a final remark, it may seem surprising that higher-order corrections to the transition amplitudes can give rise to contributions that are only of first order in the perturbation. The reason is that in protective measurement, due to the normalization of $g(t)$ [see Eq.~\eqref{eq:normal}], the strength $g(t)$ of the perturbation is linked to the duration $T$ of the perturbation (i.e., the total measurement time) and thus to the final time at which the amplitude is evaluated. For example, in the case of constant $g(t)$, the strength is precisely the inverse of $T$, $g(t)=1/T$. This interdependence between strength and duration effectively leads, for certain terms in higher-order amplitudes, to a reduction in the order of the strength parameter $1/T$.

\subsection{\label{sec:corr-constant-coupling}Pointer evolution}

We will investigate the influence of the $k_j\not=n$ contributions to Eq.~\eqref{eq:jldfjkl2} for the case of constant $g(t)$ [see Eq.~\eqref{eq:igv34763578}]. First, we look at the second-order term
\begin{align}
A^{(2)}_{n}(T) &= 
\left(-\frac{\I}{\hbar}\right)^2 \sum_{k}
\abs{\bra{n} \op{O} \ket{k}}^2 \frac{1}{T^2} \notag \\ &\quad\times \int_{-T/2}^{T/2} \D t \, \E^{\I \omega_{nk} t} \int_{-T/2}^{t} \D t' \,\E^{\I \omega_{kn} t'}.
\end{align}
The $k\not= n$ contributions are
\begin{align}
A^{(2)}_{n,k\not= n}(T) &= -\frac{\I}{\hbar^2} \sum_{k \not= n} \frac{\abs{\bra{n} \op{O} \ket{k}}^2}{\omega_{nk} T} 
\notag \\ &\quad+ \frac{1}{\hbar^2} \sum_{k \not= n} \frac{\abs{\bra{n} \op{O} \ket{k}}^2}{(\omega_{nk} T)^2}  \E^{\I \omega_{nk} T}
\notag \\ &\quad- \frac{1}{\hbar^2} \sum_{k \not= n} \frac{\abs{\bra{n} \op{O} \ket{k}}^2}{(\omega_{nk} T)^2}.
\end{align}
The first term is proportional to the second-order energy shift familiar from time-independent perturbation theory \cite{Sakurai:1994:om}
\begin{equation}
\Delta E_n^{(2)} = \frac{1}{T^2}\sum_{k \not= n} \frac{\abs{\bra{n} \op{O} \ket{k}}^2}{\hbar\omega_{nk}},
\end{equation}
the second term represents the contribution from the mixing of the other unperturbed states $\ket{m}\not=\ket{n}$, and the third term ensures wave-function normalization to second order in the perturbation. Then, from Eq.~\eqref{eq:y89ggy892} and to leading order in $1/T$, the combination of the zeroth-order, first-order, and second-order contributions to the amplitude gives
\begin{align}\label{eq:imimdfkjlkdfdfjkfqlj}
C^{(i)}_{n}(T) &\approx A^{(0)}_{n}(T)+a_iA^{(1)}_{n}(T)+a_i^2A^{(2)}_{n}(T)\notag\\
&= 1 -\frac{\I}{\hbar} a_i\bra{n} \op{O} \ket{n}  - \frac{1}{2\hbar^2} a_i^2
\bra{n} \op{O} \ket{n}^2 \notag \\ &\quad - \frac{\I}{\hbar}a_i^2\Delta E_n^{(2)}  T ,
\end{align}
where the first three terms on the right-hand side are the contributions to the pointer shift [see Eqs.~\eqref{eq:dsfdf1aa} and \eqref{eq:vcsuaadfv}]. The last term in Eq.~\eqref{eq:imimdfkjlkdfdfjkfqlj} can be thought of as the leading-order term in the expansion of the exponential $\exp\left(- \frac{\I}{\hbar} a_i^2\Delta E_n^{(2)}T \right)$. This exponential arises in the context of time-independent perturbation theory if we use the second-order perturbative approximation of the exact energy eigenvalues $\widetilde{E}(n,a_i)$ of the full Hamiltonian $\op{H}$ [compare Eq.~\eqref{eq:fnsnlk11}], 
\begin{align}\label{eq:378hgrr}
\widetilde{E}(n,a_i) &\approx E_n + \epsilon_i + \frac{1}{T}a_i \bra{n}\op{O} \ket{n} + \frac{1}{T^2} a_i^2 \sum_{k\not= n} \frac{ \abs{\bra{n}\op{O}\ket{k}}^2}{\hbar\omega_{nk}},
\end{align}
and then employ this approximation to replace the exact time-evolution term $\E^{-\I \widetilde{E}(n,a_i) T/\hbar}$ for the corresponding exact eigenstate (i.e., the state shifted from $\ket{n}$ by the perturbation) by
\begin{align}
\E^{-\I \widetilde{E}(n,a_i) T/\hbar} &\approx \E^{ -\I E_n T/\hbar}\E^{ -\I \epsilon_i T/\hbar} \E^{ -\I a_i \bra{n}\op{O} \ket{n}/\hbar} \notag \\ &\quad\times \E^{- \I a_i^2 \Delta E_n^{(2)}T/\hbar}.
\end{align}
Reintroducing the operator $\op{P}$, the last two terms on the right-hand side are equivalent to $\E^{ -\I \op{P} \bra{n}\op{O} \ket{n}/\hbar}\E^{- \I \op{P}^2 \Delta E_n^{(2)}T/\hbar}$, where $\E^{ -\I \op{P} \bra{n}\op{O} \ket{n}/\hbar}$ is the familiar pointer-shift operator. Since the argument of the term $\E^{- \I \op{P}^2 \Delta E_n^{(2)}T/\hbar}$ is proportional to $\op{P}^2$, it represents a contribution to the kinetic energy of the pointer; it induces spreading of the pointer wave packet (in addition to the free spreading) without shifting its center. 

A similar analysis may be applied to the higher-order amplitudes $A^{(\ell)}_{n}(T)$ given by Eq.~\eqref{eq:jldfjkl2} with $\ell \ge 3$. We have already discussed the term in $A^{(\ell)}_{n,\{k_j=n\}}(T)$ corresponding to setting all $k_j=n$, $1\le j \le \ell-1$, in the sum appearing in Eq.~\eqref{eq:jldfjkl2} and we have seen that this term represents a contribution to the pointer shift. Furthermore, in our analysis of higher-order corrections in Appendix~\ref{sec:high-order-contr}, we found that contributions of order $1/T$ arise from amplitudes representing a single transition (in that case, a direct transition $\ket{n}\rightarrow \ket{m}$ without intermediate virtual transitions). The analogous result holds here, but since the initial and final states are now the same, the $1/T$ terms correspond to a single virtual transition followed by a return to the initial state, i.e., the two-step transition $\ket{n}\rightarrow \ket{k} \rightarrow \ket{n}$ with $k\not =n$. Formally, these are the terms in the sum in Eq.~\eqref{eq:jldfjkl2} with a single $k_j\not=n$ and all other $k_{j'}=n$, with $j'\not=j$ and $1\le j' \le \ell-1$. They provide additional contributions to the evolution of the pointer wave packet (in the form of distortions, spreading, etc.) and are of order $1/T$ or higher. Thus, in the large-$T$ case relevant to protective measurements, they can be neglected. Additionally, as shown in Appendix~\ref{sec:high-order-contr}, they become exponentially suppressed with increasing order $\ell$. 

\section{\label{sec:eval-point-shift}Evaluation of the pointer-shift integral}

We show here that the multiple integral appearing in Eq.~\eqref{eq:1aa} is equal to $1/\ell !$. To see the idea, let us focus on the second-order double integral $\int_{-T/2}^{T/2} \D t' \, g(t') \int_{-T/2}^{t'} \D t'' \, g(t'')$. This double integral is equal to the volume $V$ under the surface given by $S(t',t'')=g(t')g(t'')$ over the region of an isosceles right triangle defined by $t' \in [-T/2, T/2]$ and $t'' \in [-T/2,t']$. Since $g(t') g(t'') = g(t'') g(t')$, the surface is symmetric about the hypotenuse of the triangle. Therefore, the volume $V$ is one-half of the volume $V_\square$ under the surface $S(t',t'')$ over the square region defined by $t' \in [-T/2, T/2]$ and $t'' \in [-T/2,T/2]$. Since $g(t)$ is normalized over $[-T/2, T/2]$, i.e., $\int_{-T/2}^{T/2} \D t \, g(t) = 1$, we have $V_\square=1$ and therefore $\int_{-T/2}^{T/2} \D t' \, g(t') \int_{-T/2}^{t'} \D t'' \, g(t'') = \frac{1}{2}$. 

We can formalize this argument and extend it to higher orders as follows (the procedure described here is similar to that used when expressing the Dyson series for the evolution operator as a time-ordered exponential). By simply exchanging the variables $t'$ and $t''$, we may write 
\begin{multline}\label{eq:1asdsda}
\int_{-T/2}^{T/2} \D t' \, g(t') \int_{-T/2}^{t'} \D t'' \, g(t'') \\ = \frac{1}{2} \biggl[ \int_{-T/2}^{T/2} \D t' \int_{-T/2}^{t'} \D t'' \, g(t') g(t'') \\+ \int_{-T/2}^{T/2} \D t'' \int_{-T/2}^{t''} \D t' \, g(t'') g(t') \biggr].
\end{multline}
In the second double integral on the right-hand side, the same region can be covered by letting $t'$ range from $-T/2$ to $T/2$ and letting $t''$ range from $t'$ to $T/2$,
\begin{multline}\label{eq:1asdsdsdda}
\int_{-T/2}^{T/2} \D t' \, g(t') \int_{-T/2}^{t'} \D t'' \, g(t'')\\ = \frac{1}{2} \biggl[ \int_{-T/2}^{T/2} \D t' \int_{-T/2}^{t'} \D t'' \, g(t') g(t'') \\+ \int_{-T/2}^{T/2} \D t' \int_{t'}^{T/2} \D t'' \, g(t'') g(t') \biggr].
\end{multline}
Thus, the double integrals can be combined into
\begin{multline}\label{eq:1asdsdsdda33}
\int_{-T/2}^{T/2} \D t' \, g(t') \int_{-T/2}^{t'} \D t'' \, g(t'') \\= \frac{1}{2} \int_{-T/2}^{T/2} \D t' \int_{-T/2}^{T/2} \D t'' \, g(t') g(t'') = \frac{1}{2},
\end{multline}
where the last step follows from the fact that $g(t)$ is normalized [see Eq.~\eqref{eq:normal}].

For arbitrary orders $\ell$, the multiple integral $\int_{-T/2}^{T/2} \D t' \, g(t') \int_{-T/2}^{t'} \D t'' \, g(t'') \cdots \int_{-T/2}^{t^{(\ell-1)}} \D t^{(\ell)} \, g(t^{(\ell)})$ is equal to the volume under the $\ell$-dimensional hypersurface in $\mathbb{R}^{\ell+1}$ given by $S(t',t'', \hdots, t^{(\ell)})=g(t')g(t'')\cdots g(t^{(\ell)})$ and we can apply the same strategy as in the $\ell=2$ case just described. Geometrically, the volume under the surface $S(t',t'', \hdots, t^{(\ell)})$ is $1/\ell !$ of the volume of the $\ell$-dimensional cube, which is $\left[\int_{-T/2}^{T/2} \D t \, g(t)\right]^\ell =1$. Hence
\begin{widetext}
\begin{equation}\label{eq:df1gfjknkjhaa}
\int_{-T/2}^{T/2} \D t' \, g(t') \int_{-T/2}^{t'} \D t'' \, g(t'') \cdots \int_{-T/2}^{t^{(\ell-1)}} \D t^{(\ell)} \, g(t^{(\ell)})  = \frac{1}{\ell !}.
\end{equation}
\end{widetext}
Instead of using this geometric argument, in analogy with Eq.~\eqref{eq:1asdsda} we could alternatively expand the multiple integral into $\ell !$ terms corresponding to the $\ell !$ possible permutations of the variables $(t',t'', \hdots, t^{(\ell)})$. Redefining the integral limits in a manner analogous to Eq.~\eqref{eq:1asdsdsdda} and using the fact that $S(t',t'', \hdots, t^{(\ell)})$ is invariant under any permutations of the variables $(t',t'', \hdots, t^{(\ell)})$, we again obtain Eq.~\eqref{eq:df1gfjknkjhaa}.


\begin{thebibliography}{23}%
\makeatletter
\providecommand \@ifxundefined [1]{%
 \@ifx{#1\undefined}
}%
\providecommand \@ifnum [1]{%
 \ifnum #1\expandafter \@firstoftwo
 \else \expandafter \@secondoftwo
 \fi
}%
\providecommand \@ifx [1]{%
 \ifx #1\expandafter \@firstoftwo
 \else \expandafter \@secondoftwo
 \fi
}%
\providecommand \natexlab [1]{#1}%
\providecommand \enquote  [1]{``#1''}%
\providecommand \bibnamefont  [1]{#1}%
\providecommand \bibfnamefont [1]{#1}%
\providecommand \citenamefont [1]{#1}%
\providecommand \href@noop [0]{\@secondoftwo}%
\providecommand \href [0]{\begingroup \@sanitize@url \@href}%
\providecommand \@href[1]{\@@startlink{#1}\@@href}%
\providecommand \@@href[1]{\endgroup#1\@@endlink}%
\providecommand \@sanitize@url [0]{\catcode `\\12\catcode `\$12\catcode
  `\&12\catcode `\#12\catcode `\^12\catcode `\_12\catcode `\%12\relax}%
\providecommand \@@startlink[1]{}%
\providecommand \@@endlink[0]{}%
\providecommand \url  [0]{\begingroup\@sanitize@url \@url }%
\providecommand \@url [1]{\endgroup\@href {#1}{\urlprefix }}%
\providecommand \urlprefix  [0]{URL }%
\providecommand \Eprint [0]{\href }%
\@ifxundefined \urlstyle {%
  \providecommand \doi  [0]{\begingroup \@sanitize@url \@doi}%
  \providecommand \@doi [1]{\endgroup \@@startlink {\doibase
  #1}doi:\discretionary {}{}{}#1\@@endlink }%
}{%
  \providecommand \doi  [0]{doi:\discretionary{}{}{}\begingroup
  \urlstyle{rm}\Url }%
}%
\providecommand \doibase [0]{http://dx.doi.org/}%
\providecommand \Doi [0]{\begingroup \@sanitize@url \@Doi }%
\providecommand \@Doi  [1]{\endgroup\@@startlink{\doibase#1}\@@Doi}%
\providecommand \@@Doi [1]{#1\@@endlink}%
\providecommand \selectlanguage [0]{\@gobble}%
\providecommand \bibinfo  [0]{\@secondoftwo}%
\providecommand \bibfield  [0]{\@secondoftwo}%
\providecommand \translation [1]{[#1]}%
\providecommand \BibitemOpen [0]{}%
\providecommand \bibitemStop [0]{}%
\providecommand \bibitemNoStop [0]{.\EOS\space}%
\providecommand \EOS [0]{\spacefactor3000\relax}%
\providecommand \BibitemShut  [1]{\csname bibitem#1\endcsname}%
\bibitem [{\citenamefont {Aharonov}\ and\ \citenamefont
  {Vaidman}(1993)}]{Aharonov:1993:qa}%
  \BibitemOpen
  \bibfield  {author} {\bibinfo {author} {\bibfnamefont {Y.}~\bibnamefont
  {Aharonov}}\ and\ \bibinfo {author} {\bibfnamefont {L.}~\bibnamefont
  {Vaidman}},\ }\href@noop {} {\bibfield  {journal} {\bibinfo  {journal} {Phys.
  Lett. A} }\textbf {\bibinfo {volume} {178}},\ \bibinfo {pages} {38}
  (\bibinfo {year} {1993})}\BibitemShut {NoStop}%
\bibitem [{\citenamefont {Aharonov}\ \emph {et~al.}(1993)\citenamefont
  {Aharonov}, \citenamefont {Anandan},\ and\ \citenamefont
  {Vaidman}}]{Aharonov:1993:jm}%
  \BibitemOpen
  \bibfield  {author} {\bibinfo {author} {\bibfnamefont {Y.}~\bibnamefont
  {Aharonov}}, \bibinfo {author} {\bibfnamefont {J.}~\bibnamefont {Anandan}}, \
  and\ \bibinfo {author} {\bibfnamefont {L.}~\bibnamefont {Vaidman}},\
  }\href@noop {} {\bibfield  {journal} {\bibinfo  {journal} {Phys. Rev. A}
  }\textbf {\bibinfo {volume} {47}},\ \bibinfo {pages} {4616} (\bibinfo {year}
  {1993})}\BibitemShut {NoStop}%
\bibitem [{\citenamefont {Anandan}\ and\ \citenamefont
  {Vaidman}(1996)}]{Aharonov:1996:fp}%
  \BibitemOpen
  \bibfield  {author} {\bibinfo {author} {\bibfnamefont {Y.~A.~J.}\
  \bibnamefont {Anandan}}\ and\ \bibinfo {author} {\bibfnamefont
  {L.}~\bibnamefont {Vaidman}},\ }\href@noop {} {\bibfield  {journal} {\bibinfo
   {journal} {Found. Phys.} }\textbf {\bibinfo {volume} {26}},\ \bibinfo
  {pages} {117} (\bibinfo {year} {1996})}\BibitemShut {NoStop}%
\bibitem [{\citenamefont {{Hari Dass}}\ and\ \citenamefont
  {Qureshi}(1999)}]{Dass:1999:az}%
  \BibitemOpen
  \bibfield  {author} {\bibinfo {author} {\bibfnamefont {N.~D.}\ \bibnamefont
  {{Hari Dass}}}\ and\ \bibinfo {author} {\bibfnamefont {T.}~\bibnamefont
  {Qureshi}},\ }\href@noop {} {\bibfield  {journal} {\bibinfo  {journal} {Phys.
  Rev. A} }\textbf {\bibinfo {volume} {59}},\ \bibinfo {pages} {2590}
  (\bibinfo {year} {1999})}\BibitemShut {NoStop}%
\bibitem [{\citenamefont {Vaidman}(2009)}]{Vaidman:2009:po}%
  \BibitemOpen
  \bibfield  {author} {\bibinfo {author} {\bibfnamefont {L.}~\bibnamefont
  {Vaidman}},\ }in\ \href@noop {} {\emph {\bibinfo {booktitle} {Compendium of
  Quantum Physics: Concepts, Experiments, History and Philosophy}}},\ \bibinfo
  {editor} {edited by\ \bibinfo {editor} {\bibfnamefont {D.}~\bibnamefont
  {Greenberger}}, \bibinfo {editor} {\bibfnamefont {K.}~\bibnamefont
  {Hentschel}}, \ and\ \bibinfo {editor} {\bibfnamefont {F.}~\bibnamefont
  {Weinert}}}\ (\bibinfo  {publisher} {Springer},\ \bibinfo {address}
  {Berlin},\ \bibinfo {year} {2009}),\ pp.\ \bibinfo {pages}
  {505--508}\BibitemShut {NoStop}%
\bibitem [{\citenamefont {Vogel}\ and\ \citenamefont
  {Risken}(1989)}]{Vogel:1989:uu}%
  \BibitemOpen
  \bibfield  {author} {\bibinfo {author} {\bibfnamefont {K.}~\bibnamefont
  {Vogel}}\ and\ \bibinfo {author} {\bibfnamefont {H.}~\bibnamefont {Risken}},\
  }\href@noop {} {\bibfield  {journal} {\bibinfo  {journal} {Phys. Rev. A}
  }\textbf {\bibinfo {volume} {40}},\ \bibinfo {pages} {2847} (\bibinfo {year}
  {1989})}\BibitemShut {NoStop}%
\bibitem [{\citenamefont {Smithey}\ \emph {et~al.}(1993)\citenamefont
  {Smithey}, \citenamefont {Beck}, \citenamefont {Raymer},\ and\ \citenamefont
  {Faridani}}]{Smithey:1993:lm}%
  \BibitemOpen
  \bibfield  {author} {\bibinfo {author} {\bibfnamefont {D.~T.}\ \bibnamefont
  {Smithey}}, \bibinfo {author} {\bibfnamefont {M.}~\bibnamefont {Beck}},
  \bibinfo {author} {\bibfnamefont {M.~G.}\ \bibnamefont {Raymer}}, \ and\
  \bibinfo {author} {\bibfnamefont {A.}~\bibnamefont {Faridani}},\ }\href@noop
  {} {\bibfield  {journal} {\bibinfo  {journal} {Phys. Rev. Lett.} }\textbf
  {\bibinfo {volume} {70}},\ \bibinfo {pages} {1244} (\bibinfo {year}
  {1993})}\BibitemShut {NoStop}%
\bibitem [{\citenamefont {Breitenbach}\ \emph {et~al.}(1997)\citenamefont
  {Breitenbach}, \citenamefont {Schiller},\ and\ \citenamefont
  {Mlynek}}]{Breitenbach:az}%
  \BibitemOpen
  \bibfield  {author} {\bibinfo {author} {\bibfnamefont {G.}~\bibnamefont
  {Breitenbach}}, \bibinfo {author} {\bibfnamefont {S.}~\bibnamefont
  {Schiller}}, \ and\ \bibinfo {author} {\bibfnamefont {J.}~\bibnamefont
  {Mlynek}},\ }\href@noop {} {\bibfield  {journal} {\bibinfo  {journal}
  {Nature (London)} }\textbf {\bibinfo {volume} {387}},\ \bibinfo {pages} {471}
  (\bibinfo {year} {1997})}\BibitemShut {NoStop}%
\bibitem [{\citenamefont {White}\ \emph {et~al.}(1999)\citenamefont {White},
  \citenamefont {James}, \citenamefont {Eberhard},\ and\ \citenamefont
  {Kwiat}}]{White:1999:az}%
  \BibitemOpen
  \bibfield  {author} {\bibinfo {author} {\bibfnamefont {A.~G.}\ \bibnamefont
  {White}}, \bibinfo {author} {\bibfnamefont {D.~F.~V.}\ \bibnamefont {James}},
  \bibinfo {author} {\bibfnamefont {P.~H.}\ \bibnamefont {Eberhard}}, \ and\
  \bibinfo {author} {\bibfnamefont {P.~G.}\ \bibnamefont {Kwiat}},\ }\href@noop
  {} {\bibfield  {journal} {\bibinfo  {journal} {Phys. Rev. Lett.} }\textbf
  {\bibinfo {volume} {83}},\ \bibinfo {pages} {3103} (\bibinfo {year}
  {1999})}\BibitemShut {NoStop}%
\bibitem [{\citenamefont {Aharonov}\ \emph {et~al.}(1988)\citenamefont
  {Aharonov}, \citenamefont {Albert},\ and\ \citenamefont
  {Vaidman}}]{Aharonov:1988:mz}%
  \BibitemOpen
  \bibfield  {author} {\bibinfo {author} {\bibfnamefont {Y.}~\bibnamefont
  {Aharonov}}, \bibinfo {author} {\bibfnamefont {D.~Z.}\ \bibnamefont
  {Albert}}, \ and\ \bibinfo {author} {\bibfnamefont {L.}~\bibnamefont
  {Vaidman}},\ }\href@noop {} {\bibfield  {journal} {\bibinfo  {journal} {Phys.
  Rev. Lett.} }\textbf {\bibinfo {volume} {60}},\ \bibinfo {pages} {1351}
  (\bibinfo {year} {1988})}\BibitemShut {NoStop}%
\bibitem [{\citenamefont {Lundeen}\ \emph {et~al.}(2011)\citenamefont
  {Lundeen}, \citenamefont {Sutherland}, \citenamefont {Patel}, \citenamefont
  {Stewart},\ and\ \citenamefont {Bamber}}]{Lundeen:2011:ii}%
  \BibitemOpen
  \bibfield  {author} {\bibinfo {author} {\bibfnamefont {J.~S.}\ \bibnamefont
  {Lundeen}}, \bibinfo {author} {\bibfnamefont {B.}~\bibnamefont {Sutherland}},
  \bibinfo {author} {\bibfnamefont {A.}~\bibnamefont {Patel}}, \bibinfo
  {author} {\bibfnamefont {C.}~\bibnamefont {Stewart}}, \ and\ \bibinfo
  {author} {\bibfnamefont {C.}~\bibnamefont {Bamber}},\ }\href@noop {}
  {\bibfield  {journal} {\bibinfo  {journal} {Nature (London)} }\textbf {\bibinfo
  {volume} {474}},\ \bibinfo {pages} {188} (\bibinfo {year}
  {2011})}\BibitemShut {NoStop}%
\bibitem [{\citenamefont {Lundeen}\ and\ \citenamefont
  {Bamber}(2012)}]{Lundeen:2012:rr}%
  \BibitemOpen
  \bibfield  {author} {\bibinfo {author} {\bibfnamefont {J.~S.}\ \bibnamefont
  {Lundeen}}\ and\ \bibinfo {author} {\bibfnamefont {C.}~\bibnamefont
  {Bamber}},\ }\href@noop {} {\bibfield  {journal} {\bibinfo  {journal} {Phys.
  Rev. Lett.} }\textbf {\bibinfo {volume} {108}},\ \bibinfo {pages} {070402}
  (\bibinfo {year} {2012})}\BibitemShut {NoStop}%
\bibitem [{\citenamefont {Born}\ and\ \citenamefont
  {Fock}(1928)}]{Born:1928:yf}%
  \BibitemOpen
  \bibfield  {author} {\bibinfo {author} {\bibfnamefont {M.}~\bibnamefont
  {Born}}\ and\ \bibinfo {author} {\bibfnamefont {V.}~\bibnamefont {Fock}},\
  }\href@noop {} {\bibfield  {journal} {\bibinfo  {journal} {Z. Phys.}
  }\textbf {\bibinfo {volume} {51}},\ \bibinfo {pages} {165} (\bibinfo {year}
  {1928})}\BibitemShut {NoStop}%
\bibitem [{\citenamefont {Fuchs}\ and\ \citenamefont
  {Peres}(1996)}]{Fuchs:1996:op}%
  \BibitemOpen
  \bibfield  {author} {\bibinfo {author} {\bibfnamefont {C.~A.}\ \bibnamefont
  {Fuchs}}\ and\ \bibinfo {author} {\bibfnamefont {A.}~\bibnamefont {Peres}},\
  }\href@noop {} {\bibfield  {journal} {\bibinfo  {journal} {Phys. Rev. A}
  }\textbf {\bibinfo {volume} {53}},\ \bibinfo {pages} {2038} (\bibinfo {year}
  {1996})}\BibitemShut {NoStop}%
\bibitem [{\citenamefont {D'Ariano}\ and\ \citenamefont
  {Yuen}(1996)}]{Ariano:1996:om}%
  \BibitemOpen
  \bibfield  {author} {\bibinfo {author} {\bibfnamefont {G.~M.}\ \bibnamefont
  {D'Ariano}}\ and\ \bibinfo {author} {\bibfnamefont {H.~P.}\ \bibnamefont
  {Yuen}},\ }\href@noop {} {\bibfield  {journal} {\bibinfo  {journal} {Phys.
  Rev. Lett.} }\textbf {\bibinfo {volume} {76}},\ \bibinfo {pages} {2832}
  (\bibinfo {year} {1996})}\BibitemShut {NoStop}%
\bibitem [{\citenamefont {Gao}(2013)}]{Gao:2013:om}%
  \BibitemOpen
  \bibfield  {author} {\bibinfo {author} {\bibfnamefont {S.}~\bibnamefont
  {Gao}},} \Eprint
  {http://arxiv.org/abs/http://philsci-archive.pitt.edu/9627}
  {http://philsci-archive.pitt.edu/9627} \BibitemShut {NoStop}%
\bibitem [{\citenamefont {Alter}\ and\ \citenamefont
  {Yamamoto}(1997)}]{Alter:1997:oo}%
  \BibitemOpen
  \bibfield  {author} {\bibinfo {author} {\bibfnamefont {O.}~\bibnamefont
  {Alter}}\ and\ \bibinfo {author} {\bibfnamefont {Y.}~\bibnamefont
  {Yamamoto}},\ }\href@noop {} {\bibfield  {journal} {\bibinfo  {journal}
  {Phys. Rev. A} }\textbf {\bibinfo {volume} {56}},\ \bibinfo {pages} {1057}
  (\bibinfo {year} {1997})}\BibitemShut {NoStop}%
\bibitem [{\citenamefont {Schlosshauer}\ and\ \citenamefont
  {Claringbold}(2014)}]{Schlosshauer:2014:tp}%
  \BibitemOpen
  \bibfield  {author} {\bibinfo {author} {\bibfnamefont {M.}~\bibnamefont
  {Schlosshauer}}\ and\ \bibinfo {author} {\bibfnamefont {T.~V.~B.}\
  \bibnamefont {Claringbold}}, in}\ \href@noop {} {\emph {\bibinfo {booktitle}
  {Protective Measurement and Quantum Reality: Towards a New Understanding of Quantum Mechanics}}},\ 
  \bibinfo {editor} {edited by\ \bibinfo {editor}
  {\bibfnamefont {S.}~\bibnamefont {Gao}}}\ (\bibinfo  {publisher} {Cambridge
  University Press},\ \bibinfo {year} {2014}),\ pp.\ \bibinfo {pages}
  {180--194}\BibitemShut {NoStop}%
\bibitem [{\citenamefont {{von Neumann}}(1955)}]{vonNeumann:1955:ii}%
  \BibitemOpen
  \bibfield  {author} {\bibinfo {author} {\bibfnamefont {J.}~\bibnamefont {{von
  Neumann}}},\ }\href@noop {} {\emph {\bibinfo {title} {Mathematical
  Foundations of Quantum Mechanics}}}\ (\bibinfo  {publisher} {Princeton
  University Press},\ \bibinfo {address} {Princeton},\ \bibinfo {year}
  {1955})\BibitemShut {NoStop}%
\bibitem [{\citenamefont {Sakurai}(1994)}]{Sakurai:1994:om}%
  \BibitemOpen
  \bibfield  {author} {\bibinfo {author} {\bibfnamefont {J.~J.}\ \bibnamefont
  {Sakurai}},\ }\href@noop {} {\emph {\bibinfo {title} {Modern Quantum
  Mechanics}}},\ \bibinfo {edition} {2nd}\ ed.\ (\bibinfo  {publisher}
  {Addison-Wesley},\ \bibinfo {address} {Reading, MA},\ \bibinfo
  {year} {1994})\BibitemShut {NoStop}%
\bibitem [{\citenamefont {Dickson}(1995)}]{Dickson:1995:lm}%
  \BibitemOpen
  \bibfield  {author} {\bibinfo {author} {\bibfnamefont {M.}~\bibnamefont
  {Dickson}},\ }\href@noop {} {\bibfield  {journal} {\bibinfo  {journal} {Philos.
  Sci.} }\textbf {\bibinfo {volume} {62}},\ \bibinfo {pages} {122} (\bibinfo
  {year} {1995})}\BibitemShut {NoStop}%
\bibitem [{\citenamefont {Anandan}(1993)}]{Anandan:1993:uu}%
  \BibitemOpen
  \bibfield  {author} {\bibinfo {author} {\bibfnamefont {J.}~\bibnamefont
  {Anandan}},\ }\href@noop {} {\bibfield  {journal} {\bibinfo  {journal}
  {Found. Phys. Lett.} }\textbf {\bibinfo {volume} {6}},\ \bibinfo {pages}
  {503} (\bibinfo {year} {1993})}\BibitemShut {NoStop}%
\bibitem [{\citenamefont {Nussinov}(1998)}]{Nussinov:1998:yy}%
  \BibitemOpen
  \bibfield  {author} {\bibinfo {author} {\bibfnamefont {S.}~\bibnamefont
  {Nussinov}},\ }\href@noop {} {\bibfield  {journal} {\bibinfo  {journal}
  {Found. Phys.} }\textbf {\bibinfo {volume} {28}},\ \bibinfo {pages} {865}
  (\bibinfo {year} {1998})}\BibitemShut {NoStop}%
\end{thebibliography}
\end{document}